\documentclass[article]{aa}
\usepackage{graphicx}
\usepackage[varg]{txfonts}
\usepackage[dvipsnames]{xcolor}
\usepackage{hyperref}
\hypersetup{pdfborder = {0 0 0},
    colorlinks = true,
    linkbordercolor = {white},
    citecolor=NavyBlue,
    urlcolor=Green
}

\newcommand{\orcidlink}[1]{\protect\href{https://orcid.org/#1}{\protect\includegraphics[width=8pt]{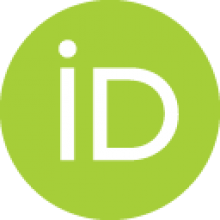}}}

\usepackage{amssymb}
\usepackage{amsmath}
\usepackage{xspace}
\usepackage{epstopdf}
\usepackage{array,multirow}
\usepackage{enumitem}
\usepackage[english]{babel}
\usepackage{mathrsfs}
\usepackage{listings}
\usepackage{blindtext}
\usepackage{lipsum} 
\usepackage{subfigure}
\usepackage[T1]{fontenc}
\usepackage[utf8]{inputenc}
\usepackage{ae,aecompl}
\usepackage{caption}
\usepackage{threeparttable}
\usepackage{booktabs}
\usepackage{longtable}
\usepackage{natbib,twoopt}
\usepackage{float}

\renewcommand{\eqref}[1]{Eq.~\ref{#1}}

\newcommand{\pvec}{\boldsymbol{p}}

\newcommand{\pH}{\pvec_\mathrm{H}}
\newcommand{\pG}{\pvec_\mathrm{G}}
\newcommand{\pHG}{\pvec^{\mathrm{H}\leftarrow\mathrm{G}}_\mathrm{G}}
\newcommand{\pCM}{\pvec^{\mathrm{H}}_\mathrm{CM}}
\newcommand{\phatG}{\widehat{\pvec}^{\mathrm{H}\leftarrow\mathrm{G}}_\mathrm{G}}

\numberwithin{equation}{section}

\makeatletter
\def\maketag@@@#1{\hbox{\m@th\normalfont\normalsize#1}}
\makeatother

\bibpunct{(}{)}{;}{a}{}{,} %% natbib format for A&A and ApJ

\makeatletter
\newcommand*\mysize{%
  \@setfontsize\mysize{5.7}{8.0}%
}
\makeatother

\makeatletter
\newcommand*\tabsize{%
  \@setfontsize\tabsize{7.}{8.0}%
}
\makeatother

\makeatletter
\newcommand\footnoteref[1]{\protected@xdef\@thefnmark{\ref{#1}}\@footnotemark}
\makeatother

\begin{document}

   \title{The $\mu$ Herculis system solved after nearly three centuries}
  \titlerunning{$\mu$ Herculis solved after three centuries}

   \author{
   Marcus L. Marcussen\inst{\ref{I1}}\orcidlink{0000-0003-2173-0689} \and 
   Mikkel N. Lund\inst{\ref{I1}}\orcidlink{0000-0001-9214-5642} \and
   Frank Grundahl\inst{\ref{I1}}\orcidlink{0000-0002-8736-1639} \and
   Daniel Huber\inst{\ref{I2}}\orcidlink{0000-0001-8832-4488} \and 
   Emil Knudstrup\inst{\ref{I1}}\orcidlink{0000-0001-7880-594X} \and 
   Adam L. Kraus\inst{\ref{I3}}\orcidlink{0000-0001-9811-568X} \and 
    Christoph Baranec\inst{\ref{I5}}\orcidlink{0000-0002-1917-9157} \and
   Pere L. Pallé \inst{\ref{I7},\ref{I8}}\orcidlink{0000-0003-3803-4823} \and
   Trent J. Dupuy \inst{\ref{I4}}\orcidlink{0000-0001-9823-1445} \and 
   Guillaume Huber \inst{\ref{I5}} \and 
   James Ou\inst{\ref{I5}}\orcidlink{0000-0002-8439-7767} \and 
   Zach Werber\inst{\ref{I5}} \and 
   Ruihan Zhang\inst{\ref{I5}} \and 
   Reed Riddle\inst{\ref{I6}}\orcidlink{0000-0002-0387-370X}}

   \offprints{MLM, \email{marcus@phys.au.dk}}          

    \institute{Stellar Astrophysics center, Department of Physics and Astronomy, Aarhus University, Ny Munkegade 120, DK-8000 Aarhus C, Denmark\label{I1}
    \and Institute for Astronomy, University of Hawai‘i, 2680 Woodlawn Drive, Honolulu, HI 96822, USA \label{I2}
    \and Department of Astronomy, University of Texas at Austin, 
    2515 Speedway, Stop C1400 Austin, TX 78712-1205, USA\label{I3}
    \and Institute for Astronomy, University of Hawai`i at M\={a}noa, Hilo, HI 96720, USA \label{I5}
    \and Instituto de Astrofisica de Canarias, 38205 La Laguna. Tenerife, Spain \label{I7}
    \and Departamento de Astrofisica. Universidad de La laguna, 38206 La Laguna. Tenerife, Spain \label{I8}
    \and Institute for Astronomy, University of Edinburgh, Royal Observatory, Blackford Hill, Edinburgh, EH9 3HJ, UK \label{I4}
    \and Division of Physics, Mathematics, and Astronomy, California Institute of Technology, Pasadena, CA 91125, USA \label{I6}
    }

   \authorrunning{Marcussen et al.}
   \date{Received xxx; accepted xxx}

  \abstract
  % context heading (optional)
  { $\mu$ Herculis is a bright, nearby quadruple system. Its brightest member, $\mu$ Her Aa, displays solar-like oscillations, establishing the system as a crucial benchmark for asteroseismology, provided that its mass can be determined independently of stellar models. }
  % aims heading (mandatory)
  { We aim to resolve the full hierarchical architecture of the system and determine precise, model-independent dynamical masses for all four components (Aa, Ab, B, and C), along with a consistent astrometric solution for the system's centre of mass. }
  % methods heading (mandatory)
  { We performed a joint fit of radial velocities, relative astrometry and absolute astrometry from \textsc{\textsc{Hipparcos}}, \textit{Gaia} DR3, and ground-based catalogues, spanning nearly three centuries. Our forward-modelling framework simultaneously constrains the Keplerian orbits of the inner Aa–Ab and B–C subsystems, the wide A--BC orbit, and the sky motion and parallax of the total centre of mass. }
  % results heading (mandatory)
  {Leveraging several complementary datasets and the decisive 2023 periastron passage of the Aa–Ab pair, we precisely determine all orbital parameters and obtain sub-percent precision on the component masses: $M_{ \rm Aa} = 1.134\pm 0.007 \,M_{\odot}$, $M_{\rm Ab} =0.2286 \pm 0.0006\,M_{\odot}$, $M_{\rm C} = 0.445 \pm 0.005\,M_{\odot}$, and $M_{\rm B} = 0.417 \pm 0.005\,M_{\odot}$. We derive a system parallax of $\varpi_{ \rm CM} = 120.069 \pm 0.089\,\mathrm{mas}$ that reconciles and improves upon the individual \textsc{Hipparcos} and \textit{Gaia} DR3 values.}
  {}
   \keywords{Quadruple stars, Binary stars, M-dwarfs, Astrometry, Radial Velocity, Asteroseismology, solar-like oscillations}

   \maketitle

%--------------------------------------------------------------------
\section{Introduction}
\label{sec:intro}

\begin{figure*}
    \centering
    \includegraphics[width=17cm]{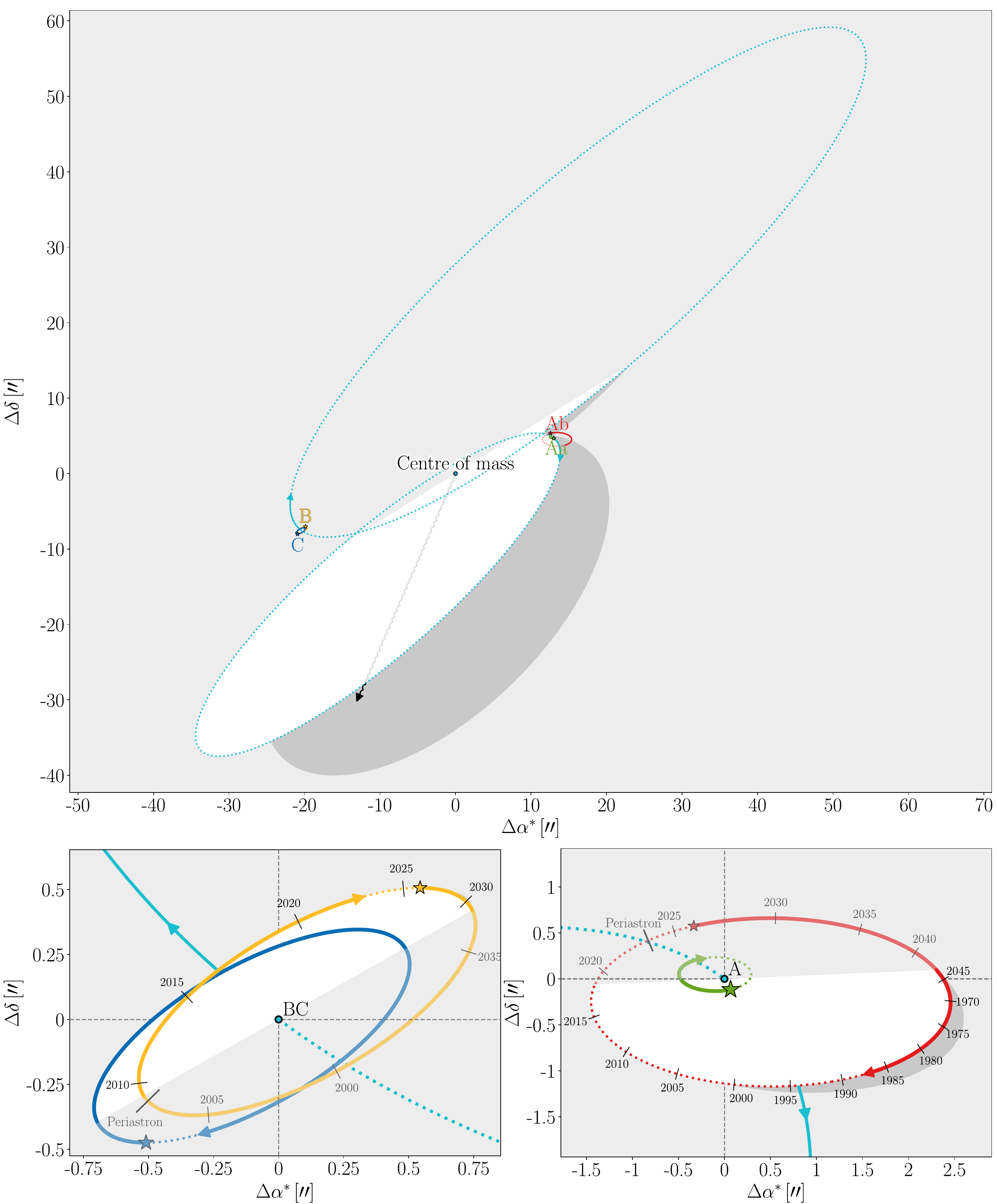}
    \caption{To-scale overview of the orbital configuration and naming scheme of the $\mu$~Her system. Components are colour-coded as Aa (green), Ab (red), B (yellow), and C (blue). The black curve traces the total sky motion of the system’s CM over a 40-year interval, combining a linear term (dashed grey line) with the annual parallax. Star symbols mark the positions on January 1 2026. Solid lines show the projected forward motion of each component over the next 40 years, and of the wide A--BC orbit over the next 400 years. To visualise the orbital inclination, segments closer to the observer are rendered in white and cast a shadow onto the sky plane; the transition between the white and grey segments marks the line of nodes. The inner A and BC subsystems are presented in greater detail in the right-hand panels.
}
    \label{fig:overview_plot}
\end{figure*}

$\mu$~Herculis ($\mu$~Her) is a nearby quadruple star system located at a distance of 8.3~pc. Its hierarchical architecture is defined by a wide A--BC configuration with a semi-major axis of ${\sim}1000$~au. The primary subsystem (A) has a semi-major axis of 20.4~au and contains the bright ($V=3.42$) G5IV subgiant $\mu$~Her~Aa (HD~161797; HIP~86974) and its M-dwarf companion, $\mu$~Her~Ab. The secondary subsystem (BC) is a tight M-dwarf pair with a semi-major axis of 11.7~au, composed of primary component $\mu$~Her~C (GJ~695~C) and secondary component $\mu$~Her~B (GJ~695~B). Figure~\ref{fig:overview_plot} shows the geometric architecture of the quadruple as derived in our analysis.

The sky position of $\mu$~Her~Aa has been tracked since antiquity: its earliest digitised entry appears in Ptolemy’s Almagest(137~AD)\citep{Lequeux2014}, and the first measurement usable in our analysis dates to 1750 \citep{Lequeux2014}. This nearly three-century record spans a substantial fraction of the wide A–BC orbit and several complete BC revolutions, while the Aa–Ab pair has been monitored for a few decades with modern techniques. Together, these data capture all geometric aspects of the system’s motion. To recover the full architecture, we assemble three complementary tracers: relative astrometry maps sky-projected orbital ellipses, radial velocities resolve the line-of-sight component, and absolute astrometry anchors the trajectory on the sky. The combination of them allows us to reconstruct all relevant orbits in a single coherent framework that naturally accounts for proper motion, orbital motion, and parallax–perspective effects, yielding precise, model-independent dynamical masses under minimal assumptions.

Earlier efforts already hinted at this possibility. \citet{Heintz1994} published orbital parameters for the Aa--Ab pair based on decades of astrometric monitoring of $\mu$~Her~Aa, and although only his final solution is available, it broadly agrees with the modern picture. Despite statements in \citet{Heintz1991} that the original measurements had been archived, extensive searches assisted by the Swarthmore Friends Historical Library\footnote{\url{https://www.swarthmore.edu/friends-historical-library}} did not recover them. More recently, tentative Aa--Ab orbits were derived by \citet{Roberts2016muHer}, using radial velocities and relative astrometry, and by \citet{Feng2021}, combining \textsc{Hipparcos} and \textit{Gaia} data. Crucially, the system has since passed through periastron, providing the curvature needed to anchor the orbital geometry and refine the component masses to high precision. The BC subsystem, discovered in 1856 \citep{Dawes1857} and observed continuously thereafter, was last characterised by \citet{Prieur2014}, who obtained a complete orbital solution but not individual masses. 

$\mu$~Her~Aa is of particular interest because it exhibits clear solar-like oscillations \citep{Grundahl2017}. With more than $120{,}000$ radial-velocity (RV) measurements made possible by the robotic nature of the Hertzsprung Stellar Observations Network Group Telescope \citep[SONG;][]{Grundahl2017,fredslund2019}, a detailed asteroseismic analysis that rivals those of the best-studied stars is underway. Combined with interferometric measurements of its radius \citep{Baines2018} and the dynamical mass derived here, $\mu$~Her~Aa will become one of the most precisely characterised stars known. The independently determined dynamical mass makes $\mu$~Her~Aa an exceptionally clean benchmark for testing and calibrating asteroseismic scaling relations and stellar models.

\section{Observations and data}
\label{sec:data}

For the Aa--Ab subsystem, we assembled five complementary datasets: (i) relative astrometry, (ii) primary-star radial velocities, (iii) absolute astrometry from \textsc{Hipparcos}, (iv) absolute astrometry from \textit{Gaia}, and (v) ground-based absolute astrometry. For the B–C subsystem and the wide A–BC pair, we relied on (i) \textit{Gaia} astrometry, (ii) published relative astrometry measurements, and (iii) a single spectroscopic measurement. To facilitate reproducibility, we compiled all observational data used in this work into a single, machine-readable archive.\footnote{Available at \url{https://www.erda.au.dk/archives/5dc20de72e83b74f490cfad4456098e4/published-archive.html}.}

\subsection{Doppler data}
To trace the reflex motion of $\mu$~Her~Aa, we used high-cadence spectra of $\mu$~Her obtained over 12 years with the spectrograph at the Hertzsprung SONG Telescope operating at Observatorio del Teide. Originally acquired for asteroseismology, these data provide exceptionally dense coverage for orbital modelling. For the present analysis, we therefore compressed each night’s time series to a single RV measurement by taking the median of all exposures. This procedure averages over solar-like p-mode oscillations and mitigates intra-night instrumental drift \citep{Chaplin2019}. Uncertainties were estimated from the scaled median absolute deviation within each night. A detector upgrade in January~2024 increased SONG’s resolving power but introduced a zero-point shift. We therefore treated pre- and post-upgrade measurements as separate instruments with independent offsets and jitter terms in the model (see Sect.~\ref{sec:methods}).

We supplemented the SONG data with archival RVs from the Hamilton and APF spectrographs at Lick Observatory \citep{rosenthal2021}, from HIRES at Keck Observatory \citep{Teklu2025}, and from CORAVEL at Haute-Provence Observatory \citep{Duquennoy1991}. We further incorporated historical measurements from \citet{Kustner1908}, \citet{Lunt1918}, \citet{Wilson1953}, \citet{Helmut1973}, \citet{Campbell1913}, \citet{Adams1923}, \citet{SpencerJones1928}, \citet{Campbell1928}, \citet{Shajn1932}, \citet{Harper1933}, and \citet{Young1939}. Table~\ref{tab:RVdata} summarises the number of RV data points used in our analysis per facility, and their uncertainties. An RV offset was applied to each unique instrument.

\begin{table}
\caption{Radial velocity data summary.}
\label{tab:RVdata}
\centering
\begin{tabular}{lll}
\toprule
Instrument & N & $\sigma\,[\mathrm{m\,s^{-1}}]$ \\
\midrule
Historical  & $18$ & $1500$ \\
CORAVEL  & $19$ & $260$ \\
Hamilton & $158$ & $2.5$ \\
HIRES & $92$ & $2.0$ \\
APF & $558$ & $1.1$ \\
$\mathrm{SONG_{old}}$ & $834$ & $2.9$ \\
$\mathrm{SONG_{new}}$ & $97$ & $2.6$ \\
\bottomrule
\end{tabular}
\tablefoot{N designates the number of observations used in our analysis, and $\sigma$ denotes the median of the formal uncertainties.}
\label{tab:RVdata}
\end{table}

Doppler coverage of the BC subsystem is significantly sparser. The sole literature value \citep{Fouque2018} is a blended measurement lacking an epoch designation, rendering it unsuitable for orbital modelling. To secure a resolved snapshot of the binary, we observed the subsystem on December 3 2025 (BJD 2461012.31) using the FIES spectrograph at the Nordic Optical Telescope \citep{NOT2010}. This observation successfully disentangled the components, yielding $v_\mathrm{C} = -11.33\pm 0.23\,\mathrm{km}\,\mathrm{s}^{-1}$ and $v_\mathrm{B} =-6.23 \pm 0.23\,\mathrm{km}\,\mathrm{s}^{-1}$.

\subsection{Relative astrometry}

Historical measurements of the relative orbits of the A--BC and BC systems were drawn from the Washington Double Star Catalogue \citep{Mason2013}. For measurements for which formal uncertainties were not originally reported, we performed a preliminary fit to characterise the residuals for each dataset, and then adopted the root mean square of these residuals as the measurement uncertainty in the final analysis.

For the Aa–Ab pair, we adopted the archival measurements compiled by \citet{Roberts2016muHer} and added two additional observations of our own. The first of these observations was obtained June 8 2023 with Keck-II and its NIRC2 AO camera, using the 2.2 $\mu$m $K_{cont}$ filter to obtain 56 short exposures ($t_{tot} = 0.48$ sec) in a subarray containing the binary. The images were reduced and two-source PSF fits were measured following \citet{Kraus2016}, where the best empirical PSF templates were adopted from the 1000 archival single-star $K_{cont}$ images that were closest in time. The astrometry was corrected for optical distortion following \citet{Service2016}. We obtained the second observation on September 17 2025 with Robo-AO-2 \cite{2024SPIE13097E..0GB}, using its natural guide star adaptive optics mode. The target was observed in the H band for a total of 5 minutes using $3\,\mathrm{ms}$ exposures. Each exposure was corrected for detector bias and sky-background before sorting by image quality. Using the best 10\% of frames, aligned to the image position of Aa, we created a final co-added image. We determined the stellar locations using the Aperture Photometry Tool.

\subsection{\textsc{Hipparcos}}
The \textsc{Hipparcos} one-dimensional abscissae epoch data are referred to as intermediate astrometric data (IAD) and exist in multiple versions and reductions, as is described in \citet{Brandt2021}. We used the latest version, extracted from the 2014 Java tool\footnote{\url{https://www.cosmos.esa.int/web/hipparcos/hipparcos-2}}. Only the bright component $\mu$~Her Aa has \textsc{Hipparcos} data, and its \textsc{Hipparcos} ID is HIP 86974.

\subsection{\textit{Gaia} DR3} 
In \textit{Gaia} DR3 \citep{GaiaDR3}, three of the four components of $\mu$~Herculis are resolved and have astrometric solutions in the \texttt{gaia\_source} table\footnote{\url{https://gea.esac.esa.int/archive/}}: $\mu$~Her Aa (Gaia DR3 4594497769766809216), $\mu$~Her~B (Gaia DR3 4594497834189213184), and $\mu$~Her~C (Gaia DR3 4594497838483177984). Our use of these catalogue parameters, and of the Gaia Observation Forecast Tool (GOST) \footnote{\url{https://gaia.esac.esa.int/gost/index.jsp}} for scan angles and epochs, is described in Sect.~\ref{sec:methods}.

\subsection{Ground-based absolute positions}
We also incorporated ground-based absolute astrometry for $\mu$~Her~Aa. Modern milliarcsecond-level positions were taken from the U.S.~Naval Observatory Bright-star Astrometric Database \citep[UBAD;][]{Munn2022}. To extend the temporal baseline, we extracted additional positions from the Harvard DASCH project \citep{Grindlay2012}, computing median co-ordinates in 10 year bins to suppress plate-to-plate systematics. Finally, we included historical catalogued positions from \citet{Boss1937}, \citet{Dyson1913}, \citet{Argelander1903}, \citet{Lequeux2014}, \citet{Fricke1963}, and \citet{Fricke1988} that help to trace the long-term motion of the system across the sky.

%--------------------------------------------------------------------

\section{Methods}
\label{sec:methods}
\subsection{Overview}
\label{subsec:model_overview}

We modelled all spectroscopic and astrometric data simultaneously in a forward-modelling framework comprising four elements: the centre-of-mass (CM) motion of the system across the sky, and the Keplerian orbits of the Aa–Ab subsystem, the B--C subsystem, and the outer A–BC system. The CM motion is described by the five standard astrometric parameters (position, proper motion, and parallax), while the orbits add the reflex displacements of each star about their respective centres of mass. All of these dynamical components and their relative strengths are shown in Fig.~\ref{fig:overview_plot}.

As is outlined in Sect.~\ref{sec:intro}, the radial velocities, relative astrometry, and absolute astrometry each constrain complementary views of the same dynamics. Our model combines all of them in a single joint fit. At each measurement epoch, our model generates the corresponding predicted value for the relevant data type. We quantified the agreement between predictions and data using a $\chi^2$ statistic that incorporates both the values and uncertainties of the measurements. From this, we built the likelihood function, which was then explored with an MCMC sampler to map the parameter space and obtain the posterior probability distribution. The full list of model parameters and their definitions is given in Table~\ref{tab:params}. Our adopted parametrisation was chosen for its close mapping to the datasets.

\begin{table}
\centering
\caption{\label{tab:params}Model parameters.}
\begin{tabular}{lll}
\toprule
Param  & Description \\
\midrule
\multicolumn{2}{c}{\textbf{Orbits}} \\
\midrule
$P$  & Orbital period. \\
$\tau$  & Time of periastron passage. \\
$e$ & Eccentricity.\\
$i$ & Inclination of the orbital axis. \\
$\Omega$  & Longitude of ascending node.\\
$\omega$  & Argument of periastron of relative orbit. \\
$a$  & Semi-major axis of the relative orbit. \\
$a_1$  & Semi-major axis of the primary’s orbit. \\
\midrule
\multicolumn{2}{c}{\textbf{Astrometry}} \\
\midrule
$\varpi$  & Parallax.  \\
$\Delta \alpha^*_{0}$ & Offset from reference position at \(t_0\) along $\alpha$. \\
$\Delta \delta_{0}$ & Offset from reference position at \(t_0\) along $\delta$. \\
$\mu_{\alpha^*}$ & Proper motion along $\alpha$. \\
$\mu_{\delta}$  & Proper motion along $\delta$. \\
\midrule
\multicolumn{2}{c}{\textbf{Other}} \\
\midrule
$f_2/f_1$ & Flux ratio \\
$\gamma_\mathrm{x}$  & Per-instrument RV zero-point offset. \\
$\sigma_\mathrm{jit,x}$ & Per-instrument RV jitter term. \\
\bottomrule
\end{tabular}
\tablefoot{$\alpha$ and $\delta$ refer to right ascension and declination co-ordinates, respectively. We adopted \(\Delta\alpha^* \equiv \Delta\alpha\cos\delta\).}
\end{table}

\begin{table*}
\caption{\label{tab:reference_solutions}Comparison of astrometric parameters for $\mu$~Her~Aa.}
\centering
\begin{tabular}{lllllll}
\toprule
Parameter & Unit & $\pH$ & $\pG$ & $\pHG$ & $\phatG$  & $\pCM$ \\
\midrule
Frame & & \textsc{Hipparcos} & \textit{Gaia} & \textit{Gaia} & \textsc{Hipparcos} & \textsc{Hipparcos} \\
$t_0$ & $\mathrm{jyear}$ & 1991.25 & 2016 & 2016 & 2016 & 1991.25 \\
$\alpha$ & $\mathrm{deg}$ & $266.615495300$ & $266.613213506$ & $266.613212289$ & $266.610790119$&$266.615667970$ \\
$\delta$ & $\mathrm{deg}$ & $27.722499170$ & $27.717275041$ & $27.717274147$ & $27.711963038$&$27.722382451$ \\
$\mu_{\alpha^*}$ & $\mathrm{mas\,yr^{-1}}$ & $-291.66 \pm 0.12$ & $-312.09 \pm 0.15$ & $-312.23 \pm 0.15$ & $-312.20 \pm 0.18$&$-324.13 \pm 0.25$ \\
$\mu_{\delta}$ & $\mathrm{mas\,yr^{-1}}$ & $-749.60 \pm 0.15$ & $-773.18 \pm 0.16$ & $-773.27 \pm 0.16$ & $-772.79 \pm 0.21$ & $-747.74 \pm 0.22$ \\
$\varpi$ & $\mathrm{mas}$ & $120.33 \pm 0.16$ & $119.86 \pm 0.15$ & $119.86 \pm 0.39$ & $120.04 \pm 0.20$ & $120.07 \pm 0.09$ \\
\midrule
$\Delta\alpha_{0}^*$ & $\mathrm{mas}$ & $0\,\pm0.09$ & $-7271.53 \pm 0.15$ & $-7275.57 \pm 0.61$ & $-14994.28 \pm 0.61$ & $550.26 \pm 8.60$ \\
$\Delta\delta_{0}$ & $\mathrm{mas}$ & $0\,\pm0.12$ & $-18806.87 \pm 0.63$ & $-18809.54 \pm 0.64$ & $-37930.08 \pm 0.64$ & $-420.19 \pm 7.55$ \\
\bottomrule
\end{tabular}
\tablefoot{$\pCM$ denotes the fitted parameters of the system's CM. $\pH$ and $\pG$ are the published \textsc{Hipparcos} and \textit{Gaia} DR3 solutions, respectively. $\pHG$ is the \textit{Gaia} DR3 solution adjusted for non-linear effects and rotated into the \textsc{Hipparcos} reference frame. $\phatG$ is the corresponding solution obtained by fitting synthetic \textit{Gaia} data points generated from our model.}
\end{table*}
\subsection{Conventions and nomenclature}

All astrometric quantities are expressed in the International Celestial Reference System (ICRS), and our model is anchored to the \textsc{Hipparcos} reference frame, meaning that we adopt $t_0 = 1991.25\,\mathrm{jyear}$, and define the position offsets, $\Delta \alpha^*_{0}$ and $\Delta \delta_{0}$, relative to the published \textsc{Hipparcos} reference position $(\alpha_H,\delta_H)$; see Table~\ref{tab:reference_solutions}.  

When an orbit is viewed through astrometry alone, it is projected onto the plane of the sky. This flattening erases the distinction between the near and far sides of the orbit, producing a \((\Omega,\omega)\!\leftrightarrow\!(\Omega+\pi,\omega+\pi)\) degeneracy. Radial velocity data provide the missing line-of-sight information, breaking this degeneracy and allowing the full three-dimensional geometry of the orbit to be determined. For purely astrometric or purely spectroscopic studies, the precise definitions of $\Omega$ and $\omega$ often matter little, since the degeneracy prevents a unique solution in any case. In our joint analysis, however, this degeneracy is finally lifted, making it crucial to state our conventions unambiguously. Unfortunately, the literature is inconsistent in how $\Omega$ and $\omega$ are defined \citep{Householder2022}, and if left unspecified, the very advantage gained by combining RVs and astrometry could be undone.  

To avoid confusion, we explicitly state our conventions, matching RV convention 2 of \citet{Householder2022}:    

\begin{itemize}
  \item Co-ordinate system: A right-handed sky-plane frame $(\hat{\mathrm{X}},\hat{\mathrm{Y}},\hat{\mathrm{Z}})$ with $\hat{\mathrm{X}}$ pointing east (along $+\alpha^*$), $\hat{\mathrm{Y}}$ pointing north (along $+\delta$), and $\hat{\mathrm{Z}}$ directed away from the observer.  
  \item Radial velocity model: The RV model uses the argument of periastron of the star being modelled, rather than that of its companion.  
  \item Ascending node: Defined as the point where the orbit crosses $Z=0$ while moving away from the observer.  
  \item Longitude of the ascending node: $\Omega$ is measured as a position angle (from north through east) to the line of nodes.  
  \item Scan angle: We adopt the \textit{Gaia} convention for the scan angle, $\psi$, measured from north to east \citep{Holl2023B}. \textsc{Hipparcos} uses a different definition, which we converted via $\psi = \pi - \psi_\mathrm{hip}$ \citep{Brandt2021}.  
  \item Orbits: The relative orbit refers to the ellipse traced by the secondary around the primary, with the primary centred at $(0,0)$. We refer to the barycentre of the Aa–Ab orbit as A, and the barycentre of the B--C orbit as BC.
  \item Subscripts: Unsubscripted quantities ($a, \omega, K, M$) refer to the relative orbit or total system values, as appropriate. Subscripts 1 and 2 refer to the primary and secondary, respectively. Using these definitions, the relative orbit is in phase with the barycentric orbit of the secondary, such that $\omega = \omega_2 = \omega_1 + \pi$.  
  \item Astrometric notation: $\pvec$ denotes the vector of the five astrometric parameters. Superscripts indicate the reference frame ($\pvec^\mathrm{H}$ for \textsc{Hipparcos}, $\pvec^\mathrm{G}$ for \textit{Gaia}, and $\pvec^{\mathrm{H}\leftarrow\mathrm{G}}$ for \textit{Gaia} rotated into the \textsc{Hipparcos} frame). Subscripts denote the entity (e.g. $\pvec_\mathrm{CM}$ for the CM). A hat, $\widehat{\pvec}$, denotes a model-implied `\textit{Gaia}-like' solution.
\end{itemize}

\subsection{Stellar mass derivation}

To express the individual component masses in terms of our fitted parameters, we start from Kepler's third law in astronomical units and years:
\begin{equation}
\label{eq:Kepler}
M = \frac{a^3}{P^2},
\end{equation}
where $M = M_1 + M_2$ is the total mass of the binary. The barycentric semi-major axes $a_1$ and $a_2$ satisfy $a = a_1 + a_2$, and the CM condition $M_1 a_1 = M_2 a_2$. These relations imply
\begin{equation}
    M_1 = M \frac{a_2}{a}, \qquad M_2 = M \frac{a_1}{a}.
\end{equation}
Substituting $M$ from Eq.~\ref{eq:Kepler} and eliminating $a_2$ via $a_2 = a - a_1$, we obtain
\begin{align}
    M_2 &= \frac{a_1 a^2}{P^2}, \qquad
    M_1 = \frac{(a - a_1)a^2}{P^2},
\end{align}
so that both component masses are written solely in terms of the fitted quantities $a$, $P$, and $a_1$.

\subsection{Forward model}
\label{subsec:forward_models}

\begin{figure*}[h!]
    \centering
    \includegraphics[width=17cm]{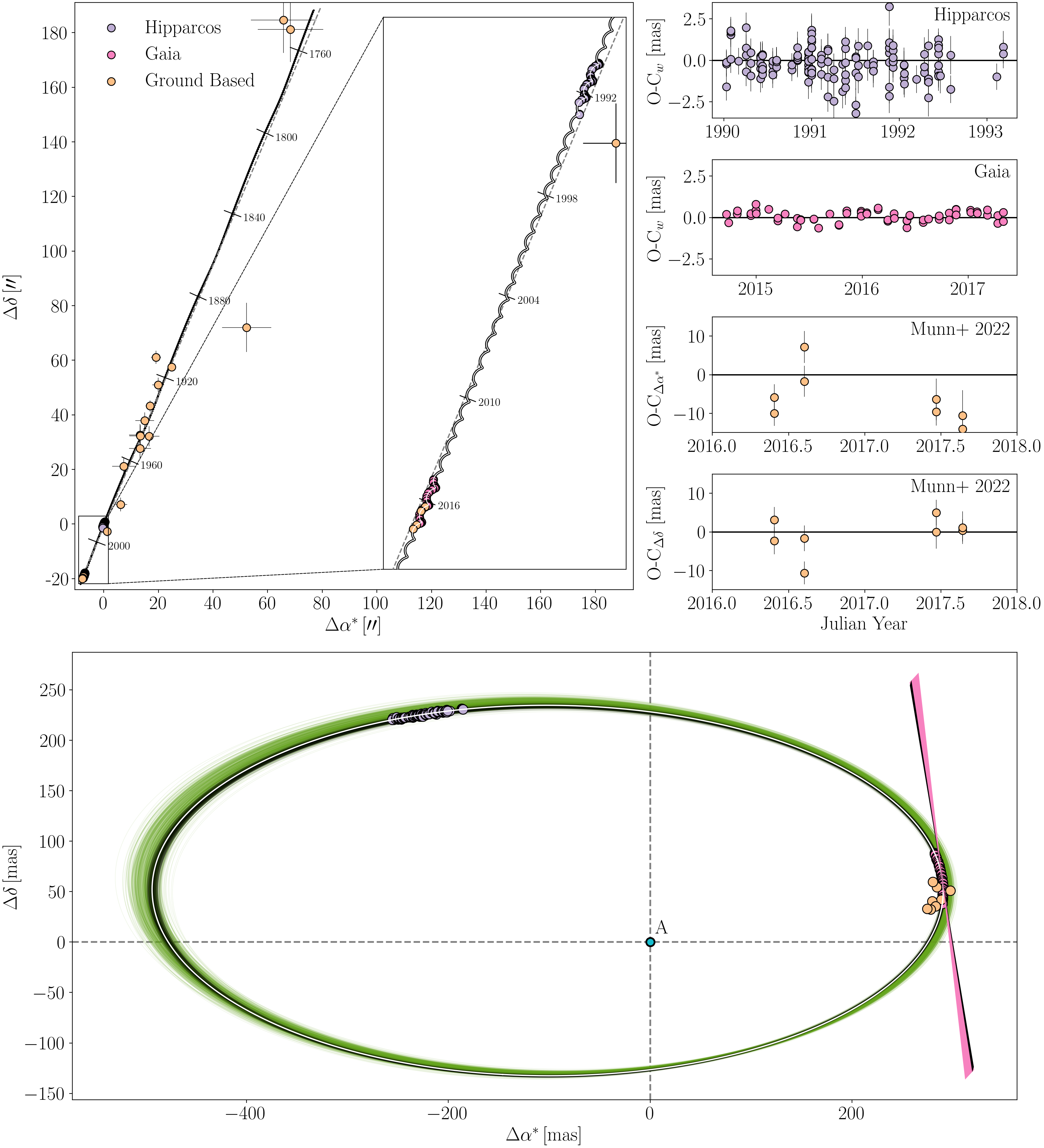}
    \caption{Sky-plane trajectory and residuals for the absolute-astrometry solution. \emph{Top:} Photocentre trajectory from our joint fit. The black curve comprises proper motion, parallax, and orbital reflex motion. The dashed grey line shows the underlying linear CM motion. \emph{Right:} Residuals. From top to bottom these panels show (I) \textsc{Hipparcos} along-scan abscissae, (II) \textit{Gaia} along-scan abscissae, (III) ground-based positions in $\alpha^*$, and (IV) ground-based positions in $\delta$. \emph{Bottom:} Orbital reflex motion of $\mu$~Her~Aa relative to the barycentre A. We display random draws from the posterior distribution of the absolute-astrometry-only solution in green and the joint solution in black. The pink cone represents the proper motion from the \textit{Gaia} solution ($3\,\sigma$ uncertainty); the black cones represent the orbital tangent vectors corresponding to the joint solutions.}
    \label{fig:skypath}
\end{figure*}

The following subsections detail the construction of the likelihood for each data type. To keep this overview concise, we defer second-order effects such as perspective acceleration, frame rotation, and the distinction between the photocentre and primary motion to Appendix~\ref{sec:appendixA}. 

We begin with the motion of the system’s CM, expressed in the tangent plane as
\begin{equation}
\begin{bmatrix}
\Delta\alpha^*_\mathrm{CM}(t) \\
\Delta\delta_\mathrm{CM}(t)
\end{bmatrix}
=
\begin{bmatrix}
\Delta\alpha^*_{0,\mathrm{CM}} \\
\Delta\delta_{0,\mathrm{CM}}
\end{bmatrix}
+
\begin{bmatrix}
\mu_{\alpha^*,\mathrm{CM}} \\
\mu_{\delta,\mathrm{CM}}
\end{bmatrix}(t-t_0)
+
\begin{bmatrix}
\Pi_{\alpha^*}(t) \\
\Pi_\delta(t)
\end{bmatrix}\varpi_\mathrm{CM},
\label{eq:cm_path}
\end{equation}
where $\Pi_{\alpha^*}(t)$ and $\Pi_\delta(t)$ are dimensionless parallax factors describing the geometric shift due to the observatory's changing vantage point. They trace the shape of the annual parallax ellipse in the local $(\alpha^*, \delta)$ basis and depend on the observatory ephemerides and the target's sky position. The overall amplitude of this apparent motion is set by the parallax $\varpi_\mathrm{CM}$. 

We model each star's sky-projected position as the sum of this CM path and two Keplerian reflex motions: one due to the orbit between the A and BC barycentres (the wide A--BC orbit), and one due to the corresponding inner orbit. To describe the orbital geometry, we use the standard Campbell elements $\{P, \tau , e, \omega, i, \Omega, a\}$. A simple approach is to first model the relative orbit and later scale it to obtain the barycentric orbits of the primary and secondary. The instantaneous linear displacement due to orbital motion in the two tangent-plane directions and along the line of sight is \citep{Leclerc2023}

\begin{align}
\label{eq:orb_rel}
\begin{bmatrix}
\Delta\alpha^*_\mathrm{orb}(t) \\
\Delta\delta_\mathrm{orb}(t) \\
\Delta Z_\mathrm{orb}(t)
\end{bmatrix}
&=
D
\begin{bmatrix}
\cos(\nu+\omega)\sin\Omega + \sin(\nu+\omega)\cos\Omega\cos i \\
\cos(\nu+\omega)\cos\Omega - \sin(\nu+\omega)\sin\Omega\cos i \\
\sin(\nu+\omega)\sin i
\end{bmatrix},
\end{align}
with
\begin{align}
D = \frac{a(1-e^2)}{1 + e\cos\nu(t)}.
\end{align}

The true anomaly, $\nu(t)$, is obtained by solving Kepler’s equation,
\begin{align}
\frac{2\pi}{P}(t-\tau) = E(t) - e\sin E(t),
\end{align}
for the eccentric anomaly, $E(t)$, and then converting via
\[
\nu(t) = 2\arctan\!\left[\sqrt{\frac{1+e}{1-e}}\,\tan\frac{E(t)}{2}\right].
\]

To obtain the secondary's position, we simply scale the relative co-ordinates of Eq.~\ref{eq:orb_rel} by $a_2/a$. The primary's position is given by the same co-ordinates scaled by $-a_1/a$, where the minus sign reflects the $180^\circ$ phase offset between the primary and relative orbits ($\omega_1 = \omega + \pi$). We are now ready to assemble the full CM-plus-orbital sky path of the Aa component:

\begin{align}
&\begin{bmatrix}
\Delta\alpha^*_\mathrm{Aa}(t) \\
\Delta\delta_\mathrm{Aa}(t)
\end{bmatrix}
= \notag \\
&
\begin{bmatrix}
\Delta\alpha^*_\mathrm{CM}(t) \\
\Delta\delta_\mathrm{CM}(t)
\end{bmatrix}-\frac{1}{\varpi_\mathrm{CM}}
\left( \frac{a_\mathrm{1,A}}{a_\mathrm{A}}\,
\begin{bmatrix}
\Delta\alpha_\mathrm{orb,A}^{*}(t) \\
\Delta\delta_\mathrm{orb,A}(t)
\end{bmatrix} +  \frac{a_\mathrm{1,A\text{-}BC}}{a_\mathrm{A\text{-}BC} }\,
\begin{bmatrix}
\Delta\alpha^*_\mathrm{orb,A\text{-}BC}(t) \\
\Delta\delta_\mathrm{orb,A\text{-}BC}(t)
\end{bmatrix}
 \right).
\label{eq:fullpath_cartesian}
\end{align}

Here, the Aa component is the primary of the Aa--Ab subsystem, and A is the primary of the A--BC subsystem; hence the orbital co-ordinates are scaled by $-a_\mathrm{1,A}/a_\mathrm{A}$ and $-a_\mathrm{1,A\text{-}BC}/a_\mathrm{A\text{-}BC}$, respectively. The factor $\varpi_\mathrm{CM}^{-1}$ converts the orbits from linear to angular size. By explicitly separating the CM motion from the orbital motion, we obtained an independent parallax for the CM, rather than relying entirely on \textit{Gaia} and \textsc{Hipparcos} catalogue solutions in which unmodelled orbital curvature can leak into the fit and bias the parallax.

To model \textsc{Hipparcos} and \textit{Gaia} data, we project the sky-plane displacement in Eq.~\ref{eq:fullpath_cartesian} onto the along-scan direction, $\psi$:
\begin{align}
w(t) = \sin\psi\,\Delta\alpha^*_\mathrm{Aa}(t) + \cos\psi\,\Delta\delta_\mathrm{Aa}(t).
\label{eq:w_projection_min}
\end{align}

The corresponding projected parallax factor is $\Pi_\psi = \Pi_{\alpha^*}\sin\psi + \Pi_\delta\cos\psi$. Finally, to model spectroscopic data, we require the radial velocity, i.e. the time derivative of the orbital $Z$-displacement, Eq.~\ref{eq:orb_rel}:
\begin{equation}
v_{x}(t) \;=\; K_x\,[\cos(\nu(t)+\omega_x) + e\cos\omega_x]\, ,
\label{eq:rv_model}
\end{equation}
where $K_x$ denotes the RV semi-amplitude (with $x=1$ for the primary and $x=2$ for the secondary),
\begin{equation}
K_x = \frac{2\pi\,a_x \sin i}{P\sqrt{1-e^2}}\, .
\end{equation}

With this forward model in place, we can evaluate the predicted observables for RV, relative astrometry, \textsc{Hipparcos}, \textit{Gaia}, and ground-based datasets and construct the joint log-likelihood:
\begin{align}
&\ln \mathcal{L}_\mathrm{tot}
= \notag \\ 
&\sum_k \ln \mathcal{L}_{\mathrm{RV},k}
+ \ln \mathcal{L}_\mathrm{RelAst}
+ \ln \mathcal{L}_\mathrm{Hip}
+ \ln \mathcal{L}_\mathrm{Gaia}
+ \ln \mathcal{L}_\mathrm{GB}\,.
\label{eq:total_like}
\end{align}
Each term is outlined below.
\subsection{Radial velocity}
\label{subsec:rv}

Spectroscopic Doppler measurements constrain the orbital parameters through RV time series. For $\mu$~Her~Aa, we have measurements, $v_{k,i}$, from multiple instruments, $k$, at epochs, $t_i$. Using Eq.~\ref{eq:rv_model}, the model-predicted RV of Aa at time $t$ is 
\begin{equation}
v_\mathrm{Aa,model}(t) \;=\; \gamma_k + v_\mathrm{1,A}(t) + v_\mathrm{1,A\text{-}BC}(t),
\end{equation}
where $\gamma_k$ is an instrument-specific velocity zero-point, and $v_\mathrm{1,A}(t)$ and $v_\mathrm{1,A\text{-}BC}(t)$ are the primary's reflex velocities induced by the inner Aa--Ab and wide A--BC orbits, respectively. For each instrument, $k$, the likelihood contribution from its RV time series is
\begin{equation}
\ln\mathcal{L}_{\mathrm{RV},k}
= -\frac{1}{2}\sum_i \left[
\frac{\left(v_{k,i}-v_{\mathrm{Aa,model}}(t_i)\right)^2}{\sigma_{k,i}^2}
+ \ln\!\left(2\pi\sigma_{k,i}^2\right)
\right],
\label{eq:like_rv}
\end{equation}
where the sum runs over $i$ measurements, and the total variance is
\[
\sigma_{k,i}^2 = \sigma_{k,i,\mathrm{obs}}^2 + \sigma_{\mathrm{jit},k}^2.
\]

Here, $\sigma_{k,i,\mathrm{obs}}$ is the formal uncertainty of measurement $i$ from instrument $k$, and $\sigma_{\mathrm{jit},k}$ is a per-instrument jitter term that captures additional noise, including stellar variability and instrumental systematics.

\subsection{Relative astrometry}
\label{subsec:rel_ast}

Relative astrometry measures the secondary's sky-projected position relative to the primary. The two observables are the angular separation, $\rho$, and the position angle of the secondary, $\theta$. From the orbital displacement in Eq.~\ref{eq:orb_rel}, the model predictions for these observables are 
\begin{align}
\rho_\mathrm{model}(t) \;&=\; 
\sqrt{\,[\varpi_\mathrm{CM}\,\Delta\alpha^*_\mathrm{orb}(t)]^2 + [\varpi_\mathrm{CM}\,\Delta\delta_\mathrm{orb}(t)]^2}, \\
\theta_\mathrm{model}(t) \;&=\; 
\mathrm{atan2}\!\left(\Delta\alpha^*_\mathrm{orb}(t),\,\Delta\delta_\mathrm{orb}(t)\right),
\end{align}
where `atan2' denotes the two-argument arctangent function. Unlike the ordinary arctangent, it uses the signs of both co-ordinate offsets to return an angle in the correct quadrant, ensuring a continuous position angle over the full $0$--$2\pi$ range. For measurements $\rho_i$ and $\theta_i$ with uncertainties $\sigma_{\rho,i}$ and $\sigma_{\theta,i}$, the log-likelihood is
\begin{equation}
\ln\mathcal{L}_{\rm RelAst} \;=\; -\frac{1}{2}\sum_i \left[
\frac{\left(\rho_i-\rho_\mathrm{model}(t_i)\right)^2}{\sigma_{\rho,i}^2}
+ \frac{\left(\theta_i-\theta_\mathrm{model}(t_i)\right)^2}{\sigma_{\theta,i}^2}
\right].
\end{equation}

Because position angles are circular quantities, the residuals, 
$\theta_i-\theta_\mathrm{model}(t_i)$, are wrapped into the interval $(-180^\circ,180^\circ]$ before squaring, ensuring continuity across the $0^\circ/360^\circ$ boundary.

\begin{figure*}[h!]
    \centering
    \includegraphics[width=17cm]{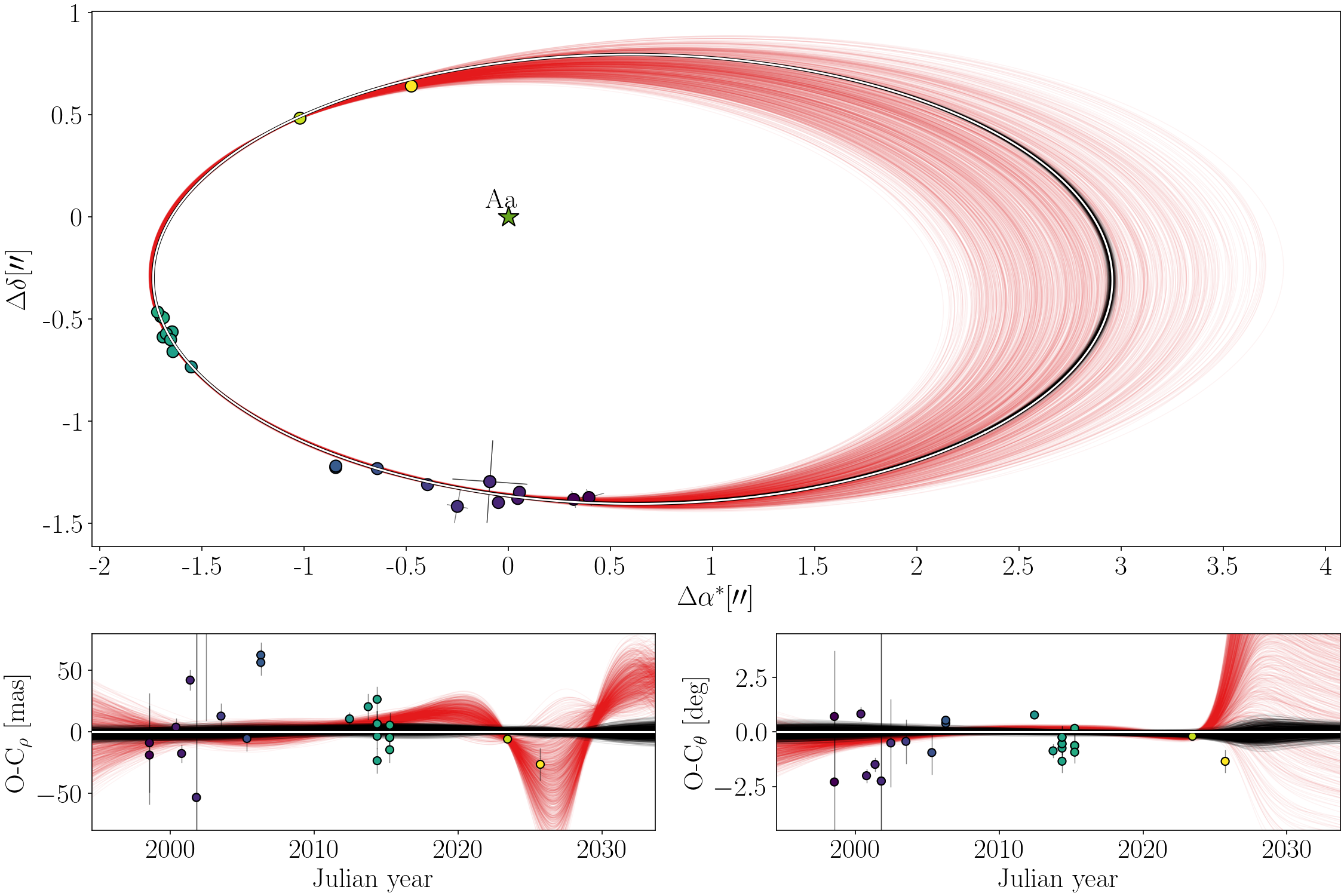}
    \caption{Relative astrometry of the $\mu$ Her A subsystem. The plot is centred on the primary $\mu$ Her Aa. Data points are colour-coded by epoch, progressing from purple to yellow. To illustrate the constraints provided by relative astrometry alone, we display 600 random draws from the corresponding posterior distribution (red traces). The solid black lines trace the orbit derived from our full joint fit, with the median solution highlighted in white. The bottom panels show residuals versus time in separation, $\rho$, and position angle, $\theta$.}
    \label{fig:astrometryA}
\end{figure*}

\subsection{Absolute astrometry}

Absolute astrometry measures sky positions in absolute co-ordinates defined relative to background stars. In this work, we used three independent sources of absolute astrometry: \textsc{Hipparcos}, Gaia, and ground-based measurements. Each is treated with a dedicated likelihood function, as described below. 

For an unresolved binary, such as $\mu$~Her Aa--Ab, the position inferred from absolute astrometry is not the position of either star individually, but corresponds to an effective image centroid (the photocentre) that lies somewhere between the two components. When the secondary is much fainter than the primary, this effective position remains close to the primary but is slightly displaced towards the companion. As the system orbits, this displacement reduces the observed amplitude of the primary's apparent barycentric motion compared to its true orbit. For clarity, when describing the modelled sky path of $\mu$~Her Aa below, we first express it in terms of the primary's barycentric position alone, neglecting this effect. In Appendix~\ref{sec:appendixA} we then describe how we model small corrections to this position that arise from the mixing $\mu$~Her Aa and $\mu$~Her Ab's light.

\begin{figure*}[h!]
    \centering
    \includegraphics[width=17cm]{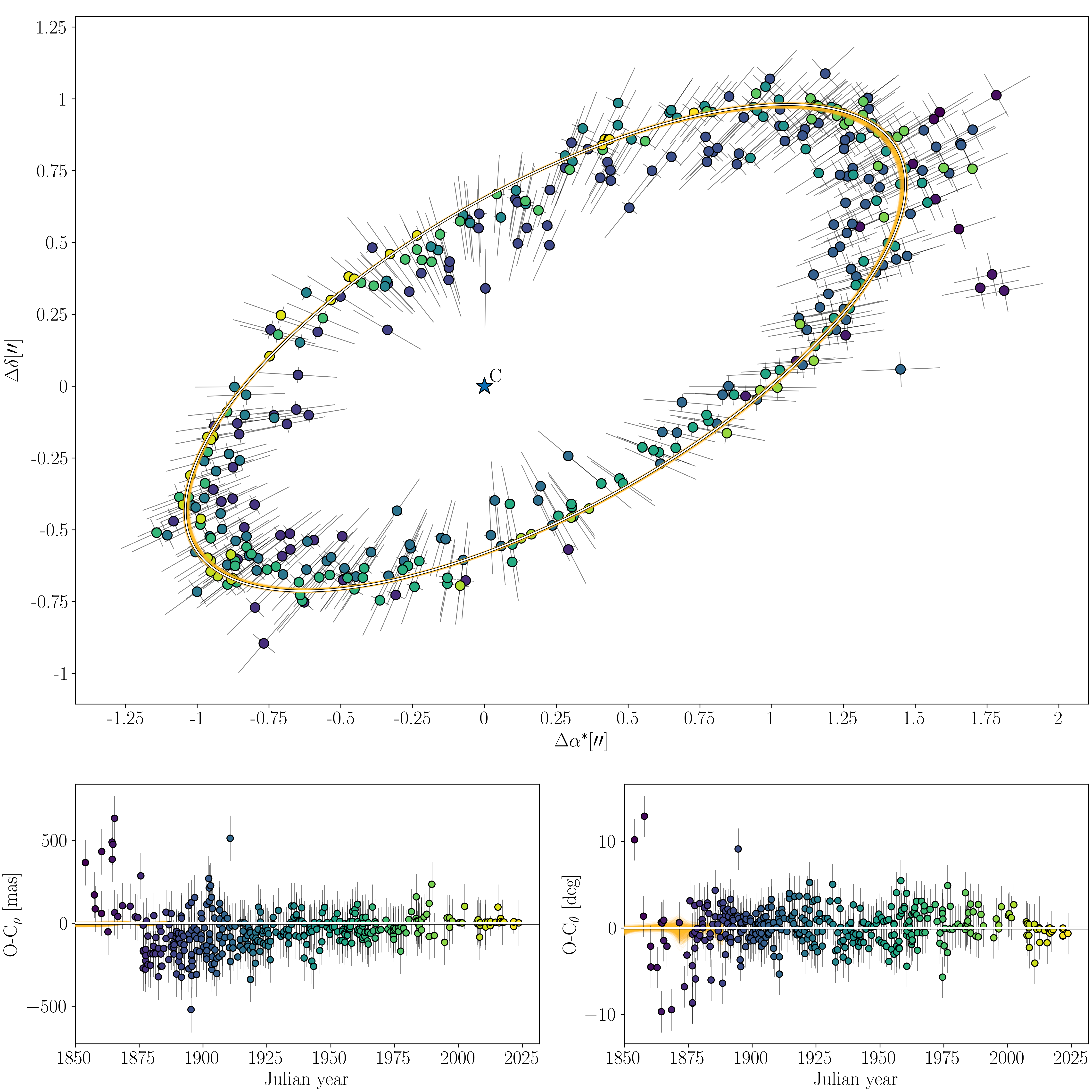}
    \caption{Relative astrometry of the BC subsystem. The plot is centred on the primary of that system, $\mu$ Her C. The format is the same as Fig.~\ref{fig:astrometryA}, but with sample orbits coloured orange.} 
    \label{fig:astrometryBC}
\end{figure*}

\subsubsection{\textsc{Hipparcos}}

For \textsc{Hipparcos}, we used IAD, which consist of the residuals of the one-dimensional along-scan abscissa measurements with respect to the published five-parameter (5p) solution. Each entry provides the epoch, $t_i$, scan angle, $\psi_i$ (after conversion to the \textit{Gaia} convention), parallax factor, $\Pi_{\psi,i}$, measured residual, $w_{\mathrm{H,res},i}$, and formal uncertainty, $\sigma_{\mathrm{H},i}$.  

Our first step towards creating the \textsc{Hipparcos} likelihood function is to use the published residuals, $w_{\mathrm{H,res},i}$ to reconstruct the abscissa measurements themselves, $w_{\mathrm{H},i}$:  

\begin{align}
w_{\mathrm{H},i} = w_{\mathrm{H,ref}}(t_i) + w_{\mathrm{H,res},i},
\end{align}
where the reference abscissa is  
\begin{align}
w_{\mathrm{H,ref}}(t) =
&\sin \psi \left[\Delta\alpha^{*}_{0,\mathrm{H}} + \mu_{\alpha^*_\mathrm{H}}(t - t_0)\right] + \notag \\
&\cos \psi \left[\Delta\delta_{\mathrm{H}} + \mu_{\delta,\mathrm{H}}(t - t_0)\right]
+ \Pi_\psi \,\varpi_\mathrm{H}. 
\end{align}  

Because our model is expressed in the \textsc{Hipparcos} reference frame and at the \textsc{Hipparcos} reference epoch, the offsets $\Delta\alpha^{*}_{0,\mathrm{H}}$ and $\Delta\delta_{0,\mathrm{H}}$ vanish by definition. The model-predicted abscissa, $w_{\mathrm{H,model}}(t)$, then follows from Eq.~\ref{eq:w_projection_min}. With the model abscissae and reconstructed measurements, we are ready to create the \textsc{\textsc{Hipparcos}} contribution to the likelihood,   

\begin{equation}
\ln \mathcal{L}_{\rm Hip} = -\tfrac{1}{2}\sum_i
\frac{(w_{\mathrm{H},i} - w_{\mathrm{H,model}}(t_i))^2}{\sigma_{\mathrm{H},i}^2}\,.
\end{equation}  

\subsubsection{\textit{Gaia} DR3}

\textit{Gaia} DR3 does not provide the underlying epoch astrometry for each source, only a linear, 5-p astrometric solution containing position, parallax, and proper motion. We therefore cannot fit the \textit{Gaia} measurements directly in the same way as the \textsc{Hipparcos} IAD. Nevertheless, the published DR3 solution still encodes strong geometric information: the sky path predicted by our model must pass close to the reported $(\alpha, \delta)$ at the DR3 reference epoch, and the local tangent to that path must align with the quoted proper-motion vector.

To exploit these constraints, we ask what \textit{Gaia} would have reported had our dynamical model been exactly correct. To answer that, we mimicked \textit{Gaia}'s measurement and fitting procedure on the model-predicted positions to obtain an astrometric solution that is directly comparable to the published one. To reconstruct the \textit{Gaia} measurements, we used the DR3 observation epochs and scan angles from the \textit{Gaia} Observation Forecast Tool. Using these epochs, we evaluated the model sky position and projected it onto \textit{Gaia}'s along-scan direction using the scan angles (Eq.~\ref{eq:w_projection_min}). This yields a set of synthetic along-scan abscissa measurements, $w_{\mathrm{G,model}}(t_i)$, which we fitted with a linear least-squares 5p model. In other words, we fit a straight-line motion to data that in reality contain slight orbital curvature, exactly as \textit{Gaia} does in its single-star solution. The fit returns the model-predicted \textit{Gaia} parameters

\begin{align}
\hat{p}^{H \leftarrow G}_{G} =
(\widehat{\Delta\alpha^{*}_{0,G}}, \widehat{\Delta\delta_{0,G}},
\widehat{\mu_{\alpha^{*}_G}}, \widehat{\mu_{\delta_G}}, \widehat{\varpi_G})\,.
\end{align}

By construction, $\hat{p}^{H \leftarrow G}_{G}$ shares the same biases as the published solution, $p^{H \leftarrow G}_{G}$, whenever our model correctly reproduces \textit{Gaia}’s observations. We built the \textit{Gaia} likelihood term from the difference between the published and fitted 5p solutions,
\begin{align}
\Delta \mathbf{p} = p^{H \leftarrow G}_{G} - \hat{p}^{H \leftarrow G}_{G},
\end{align}
with the covariance matrix, $\mathbf{C}$, derived from the published uncertainties and correlation coefficients:  
\begin{equation}
\ln \mathcal{L}_{\rm Gaia} = -\tfrac{1}{2}\,\Delta\mathbf{p}^\mathsf{T}\,\mathbf{C}^{-1}\,\Delta\mathbf{p}.
\end{equation}

\subsubsection{Ground-based astrometry}

For ground-based catalogues, the published observables are direct $(\alpha_i,\delta_i)$ positions at epochs $t_i$. We transformed them into \textsc{Hipparcos}-frame offsets, where our model operates:  
\begin{align}
\Delta\alpha^*_{\rm GB,i} &= (\alpha_i - \alpha_\mathrm{H})\cos\delta_\mathrm{H}, \notag \\
\Delta\delta_{\rm GB,i}   &= (\delta_i - \delta_\mathrm{H}).
\end{align}

Unlike \textsc{Hipparcos} or \textit{Gaia}, ground-based catalogues generally do not provide precomputed parallax factors. We therefore computed the tangent-plane factors, $\Pi_\alpha(t)$ and $\Pi_\delta(t)$, from Earth’s position with respect to the Solar System barycentre, $\mathbf{r}_\oplus(t)=(X_\oplus,Y_\oplus,Z_\oplus)$, evaluated from standard Solar System ephemerides (accessed in practice via the \texttt{Astropy} package). To evaluate these factors, we specified the target’s reference (i.e., \textsc{Hipparcos}) direction $(\alpha_\mathrm{H},\delta_\mathrm{H})$. The parallax factors are therefore
\begin{align}
\Pi_{\alpha^*}(t) &= X_\oplus(t)\sin\alpha_\mathrm{H} - Y_\oplus(t)\cos\alpha_\mathrm{H}, \\
\Pi_\delta(t) &= X_\oplus(t)\cos\alpha_\mathrm{H}\sin\delta_\mathrm{H}
+ Y_\oplus(t)\sin\alpha_\mathrm{H}\sin\delta_\mathrm{H} \notag \\
&- Z_\oplus(t)\cos\delta_\mathrm{H}.
\end{align}
With these definitions, we used Eq.~\ref{eq:fullpath_cartesian} to derive model positions at each epoch, from which we constructed the likelihood contribution from ground-based astrometry:

\begin{align}
\ln \mathcal{L}_{\rm GB} = 
&-\frac{1}{2}\sum_i
\frac{(\Delta\alpha^*_{\rm GB,i}-\Delta\alpha^*_{\rm GB,model}(t_i))^2}{\sigma^2_{\alpha^*,i}} \notag  \\
&-\frac{1}{2}\sum_i
\frac{(\Delta\delta_{\rm GB,i}-\Delta\delta_{\rm GB,model}(t_i))^2}{\sigma^2_{\delta,i}}.
\end{align}

\subsection{Sampling and priors.}
We sampled the posterior with an affine-invariant MCMC implemented in the \texttt{emcee} package \citep{Foreman-Mackey2013}, evaluating the full joint likelihood at every step. Derived quantities were recomputed at each iteration so that their uncertainties were propagated naturally through the chain. To impose uninformative priors on the orbital parameters, we sampled in the transformed variables $\cos(i)$, $\log(P)$, $\log(\sigma_\mathrm{jit})$, and $(\sqrt{e}\cos\omega,\, \sqrt{e}\sin\omega)$, which ensures uniformity over the physically relevant domains and avoids boundary artefacts.

In addition to these largely uninformative choices, we imposed a dynamical-stability prior on the wide A–BC orbit. Without such a prior, the relatively weak constraints on the outer orbit admit highly eccentric configurations in which the A and BC subsystems pass so close at periastron that the inner binaries would almost certainly be disrupted.

A definitive assessment of long-term stability would require direct $N$-body integrations of the full quadruple. Given the strong hierarchy of timescales, however, carrying out such integrations for every MCMC sample is computationally prohibitive. An alternative would be to map the stability landscape across the relevant parameters and interpolate during sampling, but resolving small stable islands in the high-dimensional parameter space at a sufficient resolution would be similarly prohibitive.

Instead, we adopted the analytical, empirically calibrated stability criterion of \citet{Mardling2001}, formulated for hierarchical triples. We treated the tighter BC pair as a single outer perturber with point mass $M_\mathrm{BC}$ on the wide orbit around the A barycentre, and evaluated the criterion for each MCMC draw. The criterion requires that the outer periastron distance, $r_p = a_\mathrm{A,BC}(1-e_\mathrm{A,BC})$, exceed the critical value $r_p^{\rm crit}$,
\begin{align}
r_p^{\rm crit} &\equiv 2.8\,a_A
\frac{(1+q)^{2/5}\,(1+e_\mathrm{A,BC})^{2/5}}{(1-e_\mathrm{A,BC})^{1/5}}
\left(1-\frac{0.3}{\pi}\Phi_\mathrm{A/A,BC}\right),
\end{align}
where $q = M_\mathrm{BC}/(M_\mathrm{Aa} + M_\mathrm{Ab})$. Because the \citet{Mardling2001} criterion was calibrated for hierarchical triples and we applied it here by approximating the BC pair as a point-mass perturber, it should be regarded as a conservative guide rather than an exact stability boundary for the full quadruple. In particular, configurations near the nominal threshold could still be stable once the full four-body dynamics are taken into account. We therefore encoded the criterion as a soft stability prior for which comfortably hierarchical solutions are essentially unaffected, while draws with very small periastron separations are progressively downweighted. Specifically, we added to our log-prior the term
\begin{equation}
  \ln p \propto -10 \ln \left[ 1 + \exp\!\left(\frac{1 - r_p/r_p^\mathrm{crit}}{0.05}\right) \right],
\end{equation}
We set the transition width to $5\%$ in $r_p/r_p^\mathrm{crit}$ to keep sampling smooth, and chose the overall strength such that clearly unstable configurations are sufficiently suppressed.

%--------------------------------------------------------------------
\section{Results}
\label{sec:results}

\begin{figure*}[h!]
    \centering
    \includegraphics[width=17cm]{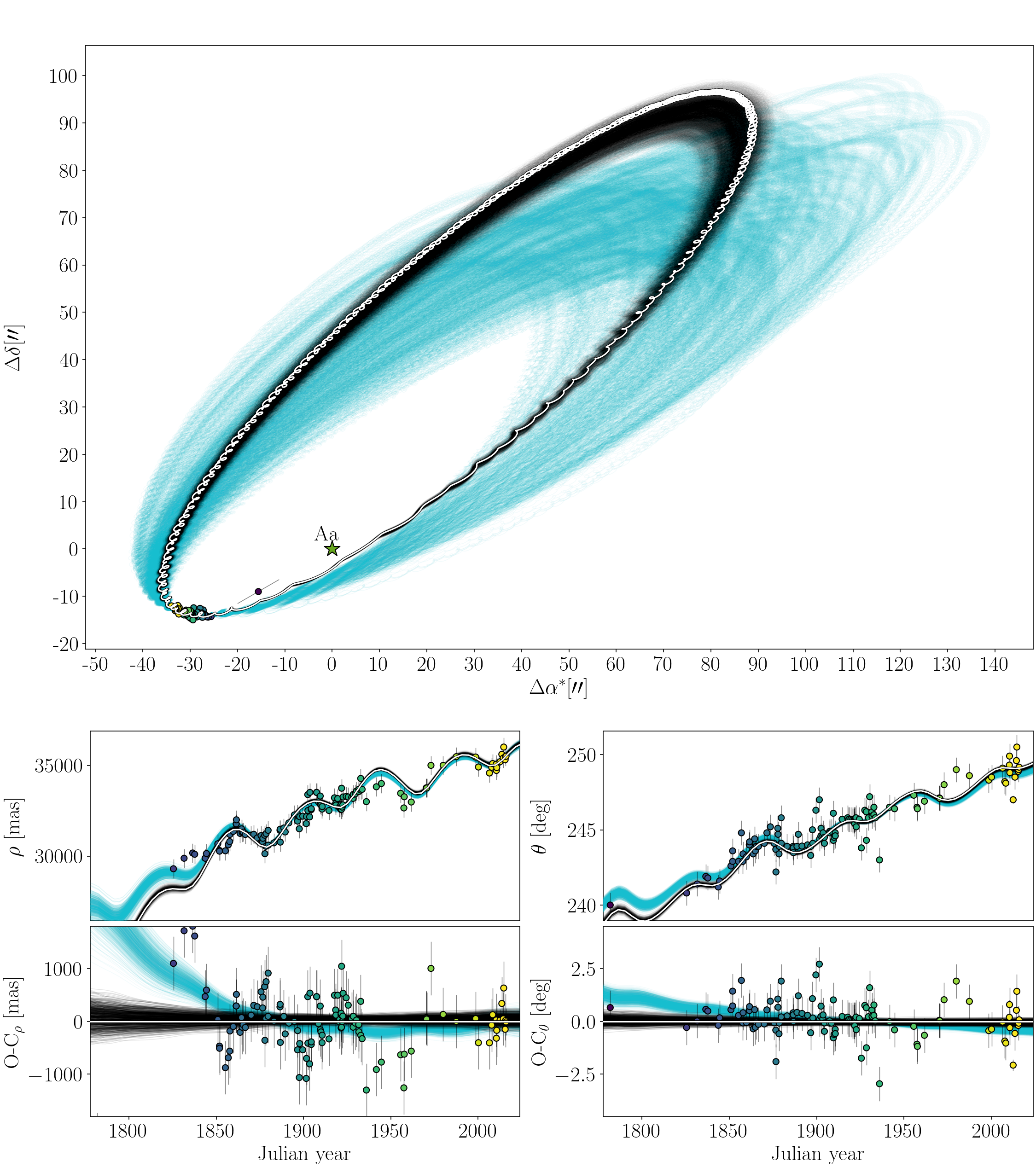}
    \caption{Relative astrometry of the outer A--BC system. The plot is centred on $\mu$ Her Aa and the format is the same as in Fig.~\ref{fig:astrometryA}. Randomly drawn orbits are coloured cyan.}
    \label{fig:astrometryOuter}
\end{figure*}

\begin{table}
\caption{\label{tab:final_params}Joint solution.}
\centering
\begin{tabular}{llll}
\toprule
Parameter & Unit & Joint Solution & $\sigma_{M_1}$ [\%] \\
\midrule
\multicolumn{4}{l}{\textbf{Astrometry}}\\ \midrule
$\varpi_{\mathrm{CM}}$ & $\mathrm{mas}$ & $120.069 \pm 0.089$ & 12.1 \\
$\mu_{\alpha^*,\mathrm{CM}}$ & $\mathrm{mas\,yr^{-1}}$ & $-324.13 \pm 0.25$ & <0.1 \\
$\mu_{\delta,\mathrm{CM}}$ & $\mathrm{mas\,yr^{-1}}$ & $-747.74 \pm 0.22$ & <0.1 \\
$\Delta \alpha^*_{\mathrm{A}}$ & $\mathrm{mas}$ & $550.26 \pm 8.60$ & <0.1 \\
$\Delta \delta_{\mathrm{A}}$ & $\mathrm{mas}$ & $-420.19 \pm 7.55$ & <0.1 \\\midrule
\multicolumn{4}{l}{ \textbf{Orbits}}\\ \midrule
$P_{\rm A}$ & $\mathrm{yr}$ & $78.987 \pm 0.288$ & 50.9 \\
$\tau_{\rm A}$ & $\mathrm{BJD}$ & $2460156.6 \pm 7.3$ & <0.1 \\
$e_{\rm A}$ &  & $0.379 \pm 0.001$ & <0.1 \\
$a_{\rm A}$ & $\mathrm{au}$ & $20.418 \pm 0.050$ & 26.6 \\
$a_{1,\rm A}$ & $\mathrm{au}$ & $3.427 \pm 0.015$ & 10.2 \\
$i_{\rm A}$ & $\mathrm{deg}$ & $62.39 \pm 0.07$ & <0.1 \\
$\omega_{\rm A}$ & $\mathrm{deg}$ & $50.303 \pm 0.224$ & <0.1 \\
$\Omega_{\rm A}$ & $\mathrm{deg}$ & $267.50 \pm 0.06$ & <0.1 \\
\midrule
$P_{\rm BC}$ & $\mathrm{yr}$ & $43.141 \pm 0.008$ & <0.1 \\
$\tau_{\rm BC}$ & $\mathrm{BJD}$ & $2454637.0 \pm 20.8$ & <0.1 \\
$e_{\rm BC}$ &  & $0.179 \pm 0.001$ & <0.1 \\
$a_{\rm BC}$ & $\mathrm{au}$ & $11.705 \pm 0.031$ & <0.1 \\
$a_{1,\rm BC}$ & $\mathrm{au}$ & $5.664 \pm 0.052$ & <0.1 \\
$i_{\rm BC}$ & $\mathrm{deg}$ & $66.10 \pm 0.09$ & <0.1 \\
$\omega_{\rm BC}$ & $\mathrm{deg}$ & $173.430 \pm 0.510$ & <0.1 \\
$\Omega_{\rm BC}$ & $\mathrm{deg}$ & $60.48 \pm 0.14$ & <0.1 \\
\midrule
$P_{\rm A,BC}$ & $\mathrm{yr}$ & $12905.46_{-608.68}^{+421.70}$ & <0.1 \\
$\tau_{\rm A,BC}$ & $\mathrm{BJD}$ & $4690500.10_{-110823.59}^{+77254.66}$ & <0.1 \\
$e_{\rm A,BC}$ &  & $0.879 \pm 0.004$ & <0.1 \\
$a_{\rm A,BC}$ & $\mathrm{au}$ & $1132_{-33}^{+25}$ & <0.1 \\
$i_{\rm A,BC}$ & $\mathrm{deg}$ & $79.58 \pm 0.21$ & <0.1 \\
$\omega_{\rm A,BC}$ & $\mathrm{deg}$ & $290.6 \pm 0.8$ & <0.1 \\
$\Omega_{\rm A,BC}$ & $\mathrm{deg}$ & $238.05 \pm 0.34$ & <0.1 \\ \midrule
\multicolumn{4}{l}{\textbf{Other}}\\ \midrule
$f_{\rm B} / f_{\rm C}$ & & $0.50 \pm 0.10$ & <0.1 \\
$\sigma_\mathrm{HIRES}$ & $\mathrm{m\,s^{-1}}$ & $3.28 \pm 0.34$ & <0.1 \\
$\gamma_\mathrm{HIRES}$ & $\mathrm{m\,s^{-1}}$ & $-16237.44 \pm 3.26$ & <0.1 \\
$\sigma_\mathrm{Hamilton}$ & $\mathrm{m\,s^{-1}}$ & $8.28 \pm 0.56$ & <0.1 \\
$\gamma_\mathrm{Hamilton}$ & $\mathrm{m\,s^{-1}}$ & $-16248.88 \pm 3.16$ & <0.1 \\
$\sigma_\mathrm{SONG_{old}}$ & $\mathrm{m\,s^{-1}}$ & $2.44 \pm 0.39$ & <0.1 \\
$\gamma_\mathrm{SONG_{old}}$ & $\mathrm{m\,s^{-1}}$ & $-16238.87 \pm 3.56$ & <0.1 \\
$\sigma_\mathrm{SONG_{new}}$ & $\mathrm{m\,s^{-1}}$ & $3.35 \pm 0.14$ & <0.1 \\
$\gamma_\mathrm{SONG_{new}}$ & $\mathrm{m\,s^{-1}}$ & $-16251.49 \pm 3.30$ & <0.1 \\
$\sigma_\mathrm{APF}$ & $\mathrm{m\,s^{-1}}$ & $3.54 \pm 0.11$ & <0.1 \\
$\gamma_\mathrm{APF}$ & $\mathrm{m\,s^{-1}}$ & $-16258.87 \pm 3.32$ & <0.1 \\
$\sigma_\mathrm{CORAVEL}$ & $\mathrm{m\,s^{-1}}$ & $0.00_{-0.00}^{+0.00}$ & <0.1 \\
$\gamma_\mathrm{CORAVEL}$ & $\mathrm{m\,s^{-1}}$ & $-16373.63 \pm 51.90$ & <0.1 \\
\midrule
\multicolumn{4}{l}{\textbf{Derived}}\\ \midrule
$M_\mathrm{Aa}$ & $\mathrm{M_\odot}$ & $1.134 \pm 0.007$ & \textemdash \\
$M_\mathrm{Ab}$ & $\mathrm{M_\odot}$ & $0.2286 \pm 0.0006$ & \textemdash \\
$M_\mathrm{C}$ &  & $0.445 \pm 0.005$ & \textemdash \\
$M_\mathrm{B}$ &  & $0.417 \pm 0.005$ & \textemdash \\
$\Phi_\mathrm{A/A-BC}$ & $^{\circ}$ & $32.59 \pm 0.24$ & \textemdash \\
$\Phi_\mathrm{BC/A-BC}$ & $^{\circ}$ & $145.58 \pm 0.24$ & \textemdash \\
$\Phi_\mathrm{A/BC}$ & $^{\circ}$ & $122.31 \pm 0.11$ & \textemdash \\
$r_p$ & $\mathrm{au}$ & $137.42 \pm 3.15$ & \textemdash \\
\bottomrule
\end{tabular}
\caption{Median and $68\%$ credible intervals on all parameters of our joint solution. The column $\sigma_{M_1}$ describes the error budget of each parameter to the mass of $\mu$~Her~Aa.}
\end{table}

Our joint analysis establishes a precise and internally coherent dynamical solution for the $\mu$~Her system. A summary of the fitted parameters is provided in Table~\ref{tab:final_params}. From these, we derived the following component masses:

\begin{align}    
&M_\mathrm{Aa} = 1.134 \pm 0.007\;M_\odot,\qquad
M_\mathrm{Ab} = 0.2286\pm 0.0006\;M_\odot, \\ \notag
&M_\mathrm{B} = 0.417\pm 0.005\;M_\odot,\qquad
M_\mathrm{C} = 0.445\pm 0.005\;M_\odot \,.\notag
\end{align}

To verify the robustness of our results, we assessed the agreement between the different observational techniques by comparing our joint solution to models constrained by individual data subsets. We began by examining the role of relative astrometry.

\subsection{Relative astrometry}
Figures~\ref{fig:astrometryA}, \ref{fig:astrometryBC}, and \ref{fig:astrometryOuter} illustrate the constraints from relative astrometry for the A, BC, and A--BC systems, respectively. The solutions derived solely from relative astrometry are consistent with our joint model, but the latter provides significantly tighter constraints on the orbital parameters, especially for A. In the case of the wide A--BC system, the measurements trace the separation between the photocentres of the A and BC binaries. Because both photocentres shift due to their respective reflex motions, the observed trajectory deviates from a simple Keplerian ellipse.

\begin{figure*}[h!]
    \centering
    \includegraphics[width=17cm]{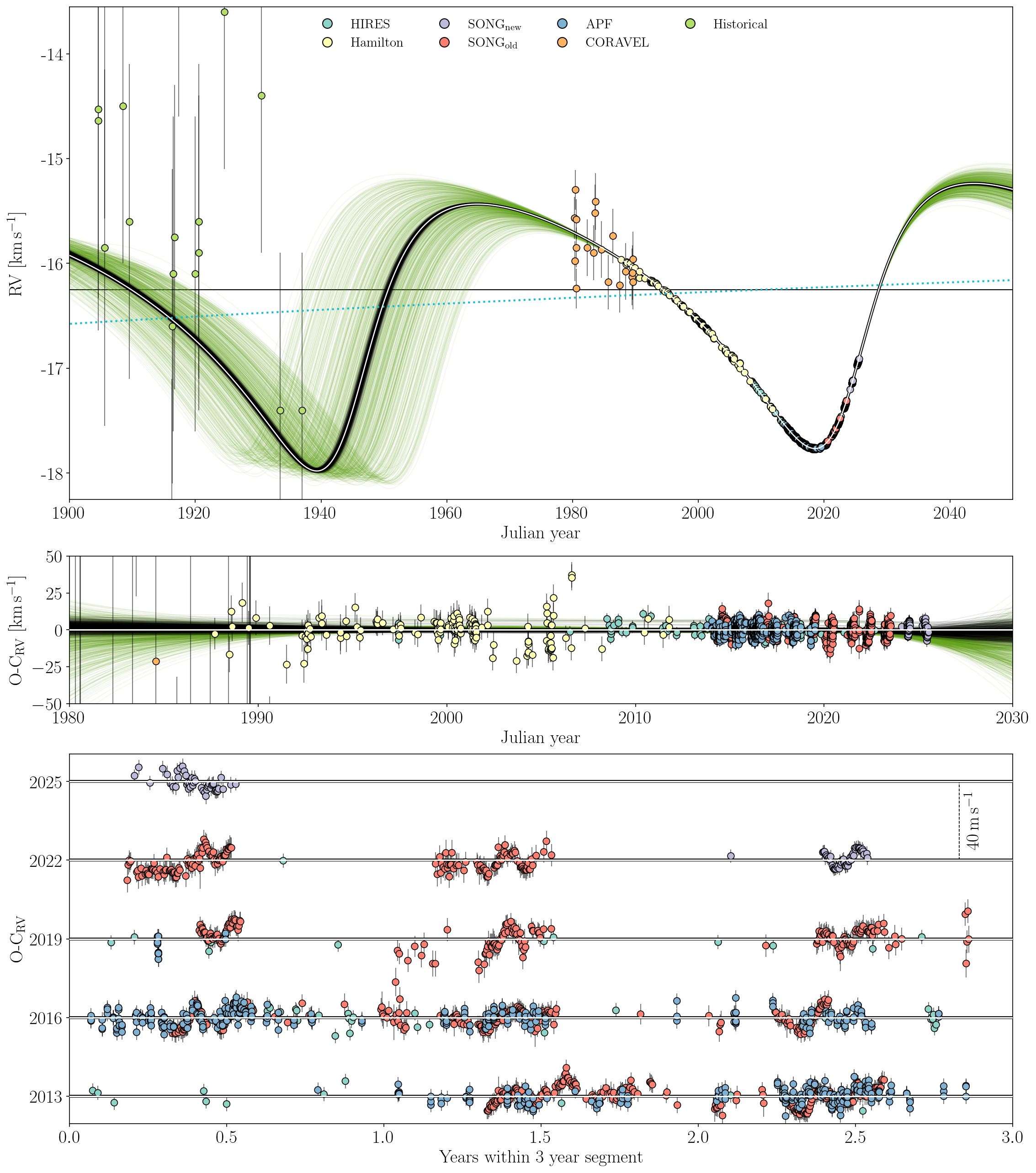}
    \caption{Radial-velocity data and model fits for $\mu$ Her Aa. \emph{Top:} Radial velocities versus time. The points are coloured by instrument (see legend). Note that SONG RVs are nightly medians. Green curves represent random draws from the posterior of the RV-only solution, while black curves depict the joint solution; the median joint model is overlaid in white. The dashed lines trace the long-term trends: the green line indicates the best-fitting linear trend for the RV-only solution, while the black line follows the reflex motion induced by the A--BC orbit in the joint fit. \emph{Middle:} Residuals relative to the median joint model. The error bars include jitter. \emph{Bottom:} Residuals in three-year segments. To visualise the dense time series, residuals are displayed in three-year segments, starting from the year indicated on the left side of the plot and vertically offset by $40\,\mathrm{ms}^{-1}$.}
    \label{fig:rv}
\end{figure*}

\subsection{Radial velocity}
Figure~\ref{fig:rv} summarises the $\mu$~Her~Aa RV data and models. Considered in isolation, the RVs leave the orbital period, $P_\mathrm{A}$, periastron epoch, $\tau_\mathrm{A}$, and semi-amplitude, $K_\mathrm{1,A}$, only weakly constrained, although the shape parameters, $e_\mathrm{A}$ and $\omega_\mathrm{A}$, are relatively well determined from the velocity curve. The constraints strengthen once the angular scale of the orbit is set by relative astrometry. In the joint fit, the RVs fix the primary's barycentric semi-major axis, $a_\mathrm{1,A}$, in addition to refining $e_\mathrm{A}$ and $\omega_\mathrm{A}$. For the RV-only solution, we allowed a linear RV drift as a function of time to absorb the contribution from the A--BC system. 

In the highest-precision APF and SONG data, the RV residuals show a coherent, low-amplitude structure that we ascribe to rotation-modulated stellar activity; a detailed analysis will be presented in Kjeldsen (in prep.). We modelled this with per-instrument jitter terms at the level of a few meters per second, which absorbed the activity signal without materially biasing the orbital elements. Where the APF and SONG time series overlap, the residual patterns are similar, supporting an astrophysical rather than an instrumental origin (see Fig.~\ref{fig:rv}).

For the BC subsystem, our single spectroscopic observation of $\mu$~Her~BC exhibits asymmetric absorption lines. These asymmetries indicate that, at the epoch of observation, the brighter C component was approaching along the line of sight, while the B component was receding. To disentangle the contributions from B and C and recover individual radial velocities, we computed broadening functions \citep{Rucinski1999} based on model atmospheres from \citet{Kurucz1979}, and fitted two Gaussian profiles to the resulting distribution. We adopted uncertainties from the diagonal elements of the covariance matrix from the least-squares minimisation. The broadening function and extracted velocities are displayed in Fig.~\ref{fig:BF}. 

\subsection{Absolute astrometry}
The absolute astrometry data trace the sky-plane trajectory of A’s photocentre. This composite path encodes the combined effects of proper motion, parallax, and the inner and outer orbits, and therefore constrains the full geometry of the model. In practice, the most direct contribution from absolute astrometry is the system parallax. Apart from this role, the absolute astrometry data is broadly analogous to that of the Aa radial velocities: it constrains the orientation of the Aa--Ab orbit on the sky and sets the scale of the primary's barycentric semi-major axis, $a_{1,A}$. Much of this constraining power comes from the \textit{Gaia} DR3 proper motion, which effectively measures the local tangent of $\mu$~Her~Aa's trajectory.

\subsubsection{Parallax}
Our joint analysis yields a system parallax of $\varpi_\mathrm{CM} = 120.069\pm0.089~\mathrm{mas}$. This result reconciles the independent measurements of Aa from \textsc{Hipparcos} with the \textit{Gaia} parallaxes for components Aa, B, and C. Moreover, the published \textit{Gaia} DR3 5p solution (transformed into the \textsc{Hipparcos} frame, $\pHG$) agrees with the solution implied by our dynamical model ($\phatG$), demonstrating their mutual consistency (see Table~\ref{tab:reference_solutions}).

\subsubsection{Orbital parameters}
The top left panel of Fig.~\ref{fig:skypath} displays the total projected sky path, while the bottom panel subtracts the parallax, proper motion, and outer orbit to isolate the reflex motion of Aa. To test the internal consistency of the absolute astrometry, we performed two validation fits and compared them to our joint solution. First, we solved the Aa--Ab orbit while omitting absolute astrometry data entirely, yielding the result $M_\mathrm{Aa} = 1.134 \pm 0.009\,\mathrm{M_\odot}$. Second, we omitted radial velocities and let absolute astrometry alone determine $a_{1,A}$. The resulting solution, also shown in Fig.~\ref{fig:skypath}, yields $M_\mathrm{Aa} = 1.149 \pm 0.011\,M_\odot$.

\subsubsection{BC component}
For components B and C, the available absolute astrometry is limited to their \textit{Gaia} DR3 catalogue solutions. After subtracting both the CM proper motion and the contribution from the wide A--BC orbit, the residual proper motions represent the instantaneous sky-plane orbital motion of B and C about their shared barycentre. This information, in turn, allows us to determine the individual masses of the BC subsystem.

\subsection{Final parameter estimates}
To understand what limits the precision on $M_{\rm Aa}$, we decomposed its variance into contributions from each fitted parameter. Because many parameters are correlated in the joint posterior, this decomposition must respect their covariance structure, and we therefore worked directly with the MCMC samples. From these, we computed the covariance matrix, $C_{xx}$, of the model parameters and the vector of cross-covariances, $C_{xM_\mathrm{Aa}}$, between each parameter and the derived mass. The linear relation

\begin{equation}
\boldsymbol{\beta} = C_{xx}^{-1} C_{xM_\mathrm{Aa}}
\end{equation}
then gives an effective gradient: it describes how small changes in each parameter translate into changes in $M_{\rm Aa}$ near the posterior mean. The contribution of parameter $i$ to the total variance of the mass is
\begin{equation}
S_i = \beta_i \,(C_{xM_\mathrm{Aa}})_i .
\end{equation}
Table~\ref{tab:final_params} lists these $S_i$ as normalised percentages, providing an error budget for the dynamical mass of Aa. This exercise reveals that $P_A$ is the main driver of uncertainty to $M_\mathrm{Aa}$, followed by $a_A$. This result is to be expected, since our data do not cover a full orbit of the A subsystem, in either RV or relative astrometry. Corner plots showing the posterior correlations among the orbital and astrometric parameters are provided in Figs.~\ref{fig:corner_Astrometry}, \ref{fig:corner_inner}, \ref{fig:corner_BC}, and \ref{fig:corner_outer}.

\subsection{Mutual inclinations and spin-orbit angles}

By combining radial velocities with absolute and relative astrometry, we lifted the degeneracy between the ascending and descending nodes for all orbits in the system. With the full three-dimensional geometry accessible, we could determine the mutual inclinations, $\Phi$, between the orbital planes and thereby map the spatial architecture of the quadruple. The mutual inclination between any two orbits follows from
\[
\cos \Phi = \cos i_1 \cos i_2 + \sin i_1 \sin i_2 \cos(\Omega_1 - \Omega_2),
\]
from which we inferred $\Phi_{A/\mathrm{A\text{-}BC}} = 44.94\pm3.22^\circ$ for the inner A subsystem relative to the outer A–BC orbit, $\Phi_{\mathrm{BC/A\text{-}BC}} = 140.00\pm2.24^\circ$ for the BC subsystem relative to A–BC, and $\Phi_{\mathrm{A/BC}} = 122.27\pm0.12^\circ$ between the two inner binaries; see also Table~\ref{tab:final_params}.

Additionally, the stellar inclination of $\mu$~Her~Aa was measured from asteroseismology by \citet{Grundahl2017} as $i_\star = 63^{+9}_{-10}{}^\circ$. This value is fully consistent with our orbital inclination for the Aa--Ab pair, $i_A = 62.53 \pm 0.07^\circ$, and therefore provides no evidence of a misalignment between the stellar spin axis and the orbital axis, although an azimuthal misalignment remains possible. Empirical constraints on spin--orbit alignment for binaries in the $3$--$100\,\mathrm{au}$ range \citep{Justesen2020} remain sparse, but by analogy with trends observed at smaller separations \citep{Marcussen2024}, spin-orbit alignment in $\mu$~Her~A would not be unexpected.

\section{Discussion}
\label{sec:discussion}

\subsection{Formation and dynamical history}

The present-day orbital architecture of $\mu$\,Her is markedly non-coplanar. The inner Aa--Ab orbit is inclined by ${\sim}\,45^\circ$ with respect to the wide A--BC orbit, while the BC subsystem is retrograde at ${\sim}\,140^\circ$. With an outer semi-major axis of ${\sim}\,1100\,\mathrm{au}$, the system falls between the proposed formation pathways of core fragmentation and disc fragmentation, as is discussed by \citet{Offner2023}. A more eccentric and misaligned architecture is expected from core fragmentation, whereas a more circular and aligned system is characteristic of disc fragmentation. The observed geometry naively suggests the former pathway. However, the close periastron passages in the wide orbit are expected to drive secular effects, such as chaotic eccentric Kozai--Lidov cycles \citep{Naoz2016}, which can substantially reshape an initially simple configuration and partially erase its formation signature. 

An additional clue about the system's history is provided by the inferred stellar inclination of Aa, which is consistent with spin--orbit alignment in the A subsystem. This suggests that the inner Aa–Ab pair has formed in a shared circumbinary disc, and that the system has likely not experienced strong inclination--eccentricity cycling since formation. Its present orientation may thus still be close to primordial. By contrast, the BC pair is both less massive and more compact, and therefore carries a smaller share of the system's total angular momentum. In the presence of the highly eccentric wide orbit, long-term secular torques can more readily reorient such a low--angular--momentum inner binary, potentially driving it to the high inclination we observe today.

\subsection{$\mu$ Herculis as a benchmark system}

The precise dynamical mass derived here establishes $\mu$~Her~Aa as a premier benchmark for stellar physics. By setting the fundamental scale of the system independent of stellar evolution models, we provide a rigorous testbed for validating asteroseismic scaling relations and examining the internal physics of subgiants. 

\begin{figure}
    \resizebox{\hsize}{!}{\includegraphics{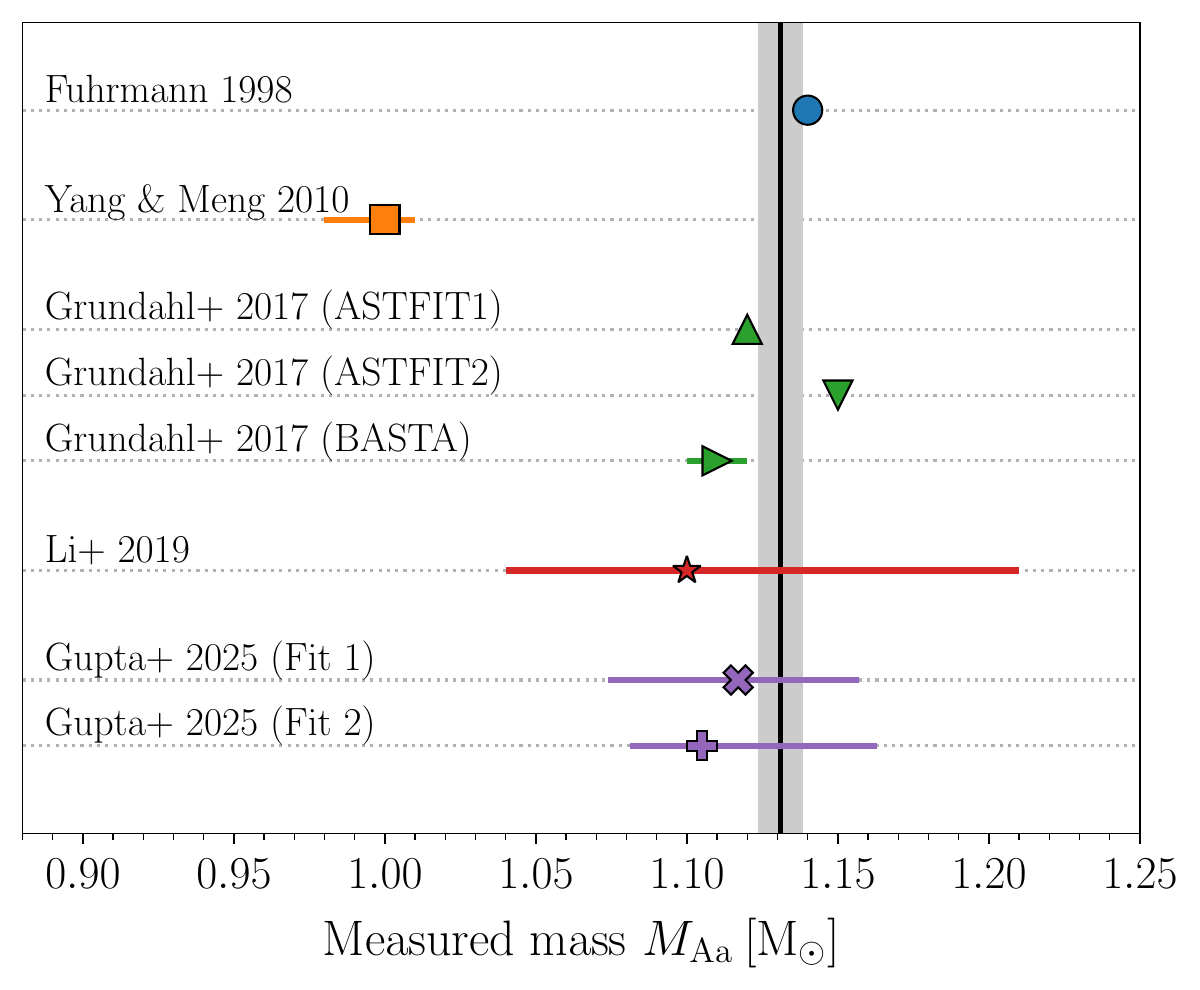}}
    \caption{Comparison of the primary mass, $M_{Aa}$, derived in this work (vertical grey band, showing the $1\sigma$ credibility interval) against literature values. The markers represent estimates from non-seismic spectroscopy \citep{Fuhrmann1998}, stellar modelling \citep{Yang2010, Li2019}, and asteroseismic analyses \citep{Grundahl2017, Gupta2025arXiv}.}
    \label{fig:mass_compare}
\end{figure}

Prior to this work, estimates of the primary mass relied on spectroscopic calibrations or model-dependent asteroseismic scaling, leading to a significant spread in values. Figure~\ref{fig:mass_compare} compares our model-independent dynamical mass against a representative selection of these literature values.

Our dynamical mass for $\mu$~Her~Aa, $1.134\,M_{\odot}$, falls slightly below the early non-seismic spectroscopic estimate by \citet{Fuhrmann1998} ($1.14\,M_{\odot}$), while remaining at the upper end of the masses inferred from asteroseismic and evolutionary modelling. In the first modelling efforts based on the initial oscillation detections of \citet{Bonanno2008}, \citet{Yang2010} reported a best-fit mass of $1.00\,M_{\odot}$, noting that solutions of up to $1.10\,M_{\odot}$ could be obtained under alternative assumptions about the input physics.

With the advent of higher-precision seismic constraints, the inferred masses moved upwards but remained below the early spectroscopic value. The grid modelling by \citet{Li2019} yielded $1.10^{+0.06}_{-0.02}\,M_{\odot}$, and the solutions presented by \citet{Grundahl2017} (ASTFIT, BASTA) cluster close to $1.12\,M_{\odot}$. Most recently, analyses incorporating TESS data and updated glitch modelling by \citet{Lund2025} and \citet{Gupta2025arXiv} have remained in this same regime, with \citet{Gupta2025arXiv} deriving $1.105^{+0.058}_{-0.024}\,M_{\odot}$.

As is illustrated in Fig.~\ref{fig:mass_compare}, our dynamical mass therefore provides an independent, geometry-driven constraint that is broadly compatible with the seismic consensus, while selecting its high-mass end. Any successful asteroseismic or evolutionary model of $\mu$~Her~Aa must reproduce both the detailed oscillation spectrum and this dynamical mass.

Beyond its role in calibrating asteroseismology, the $\mu$~Her system also offers an opportunity to test models of low-mass stars. Existing spectroscopic and asteroseismic analyses already indicate that $\mu$~Her~Aa is an old, metal-rich subgiant \citep[e.g.][]{Fuhrmann1998,Li2019,Gupta2025arXiv}, and the dedicated modelling of the SONG asteroseismic data will deliver precise constraints on its age and heavy-element abundance. Assuming coeval formation, these values can be adopted for Ab, B, and C; combined with our dynamical masses, the three M dwarfs then constitute a small sequence of benchmark stars at fixed age and supersolar metallicity. A dedicated comparison of their radii, luminosities, and spectra with low-mass evolutionary and atmospheric grids would provide a stringent test of M-dwarf models, but we leave such an analysis to future work.

\subsection{Future prospects}

While the current solution is robust, several avenues remain to further tighten the constraints on this system. The release of \textit{Gaia} DR4 will provide time-series epoch astrometry, enabling a direct fit to the \textit{Gaia} measurements that should further improve the precision of orbital elements and the system parallax. At present, the error budget of $M_{Aa}$ is dominated by the relatively small number of astrometric measurements of the Aa--Ab separation and their modest temporal coverage. Additional high-precision measurements taken over a few years would therefore substantially improve the accuracy of the masses in the A subsystem.

\begin{acknowledgements}
This publication includes observations made with the SONG network of telescopes operated by Aarhus University, Instituto de Astrofísica de Canarias, the National Astronomical Observatories of China, the University of Southern Queensland and New Mexico State University.

MNL acknowledges support from the ESA PRODEX programme (PEA 4000142995). 

This research has made use of the SIMBAD database \citep{wegner2000}, operated at CDS, Strasbourg, France.

This research has made use of the Washington Double Star Catalog maintained at the U.S. Naval Observatory.

This work presents results from the European Space Agency (ESA) space mission \textit{Gaia}. \textit{Gaia} data are being processed by the \textit{Gaia} Data Processing and Analysis Consortium (DPAC). Funding for the DPAC is provided by national institutions, in particular the institutions participating in the \textit{Gaia} MultiLateral Agreement (MLA). The \textit{Gaia} mission website is \url{https://www.cosmos.esa.int/gaia}. The \textit{Gaia} archive website is \url{https://archives.esac.esa.int/gaia}.

This research made use of the SONG database SODA (\url{https://soda.phys.au.dk/}), operated and maintained at Aarhus University, DK.      

The Robo-AO-2 system is supported by the National Science Foundation under Grant Nos. AST-1712014 and AST-2509941, the State of Hawaii Capital Improvement Projects, the Mt. Cuba Astronomical Foundation, and by a gift from the Lumb Family. Support for the infrared camera for Robo-AO and Robo-AO-2 was provided by the Mt. Cuba Astronomical Foundation and through the National Science Foundation under Grant No. AST-1106391

Based on observations made with the Nordic Optical Telescope,
owned in collaboration by the University of Turku and Aarhus University,
and operated jointly by Aarhus University, the University of Turku, and
the University of Oslo, representing Denmark, Finland and Norway, the University
of Iceland and Stockholm University at the Observatorio del Roque de los
Muchachos, La Palma, Spain, of the Instituto de Astrofisica de Canarias.

We acknowledge the use of the following Python-based software modules:  \texttt{Astropy} \citep{Astropy}, \texttt{scikit-image} \citep{scikit-image}
\end{acknowledgements}

\bibliographystyle{aa} 
\bibliography{biblio}

\begin{appendix}
\section{Additional effects}
\label{sec:appendixA}

In the foregoing description, we intentionally omitted several small effects. In nearby, high–proper-motion systems such as $\mu$ Herculis, however, these effects are no longer negligible at the precision we seek. Here, we introduce the additional terms needed to relax those assumptions, explain how each is incorporated into the model, and quantify its contribution. Full derivations are given in the cited references.

\subsection{Photocentre versus\ primary}
When a binary system is not fully resolved, the measured position on the sky corresponds not to either star individually but to a weighted blend of their light, the photocentre. This distinction matters because the secondary’s flux pulls the centroid of the system away from the true position of the primary. In effect, the apparent orbital motion of the primary is diluted by the contribution from the companion.

We capture this dilution through the factor $\beta$, which links the photocentre displacement $(\Delta\alpha^*_{\mathrm{orb,0}}, \Delta\delta_{\mathrm{orb,0}})$ to the relative orbital co-ordinates $(\Delta\alpha^*_{\mathrm{orb}}, \Delta\delta_{\mathrm{orb}})$ as
\begin{align}
\begin{bmatrix}
\Delta\alpha^*_{\mathrm{orb,0}}(t) \\ 
\Delta\delta_{\mathrm{orb,0}}(t)
\end{bmatrix}
=
\left(\beta - \frac{a_1}{a}\right)
\begin{bmatrix}
\Delta\alpha^*_{\mathrm{orb}}(t) \\ 
\Delta\delta_{\mathrm{orb}}(t)
\end{bmatrix}.
\end{align}

In the limit of small separations, where the secondary cannot be distinguished in the point-spread function, $\beta$ takes its maximum value:
\begin{align}
\beta = \frac{f_2/f_1}{1+f_2/f_1},
\end{align}
with $f_1$ and $f_2$ denoting the fluxes of the primary and secondary. At very large separations, $\beta$ tends to zero. For \textit{Gaia}, the transition between these regimes occurs at angular separations of roughly $9\,\mathrm{mas}$ (fully blended) and $270\,\mathrm{mas}$ (fully separated; \citealt{Lindegren2022}). At intermediate separations, $\beta$ can be modelled using Eq.~12 of \citet{Lindegren2022}.

During the \textit{Gaia} DR3 observations, $\mu$~Her Aa--Ab had a mean separation of ${\sim}1700\,\mathrm{mas}$, well beyond \textit{Gaia}’s nominal blending limit. Nonetheless, the one-dimensional nature of \textit{Gaia}’s along-scan measurements means that the effective separation scales by how much the line between the primary and secondary align with scan angle: $\Delta\eta =\rho \cos(\psi - \theta)$. Thus, even in such a wide system, a small subset of the \textit{Gaia} measurements still registers a small photocentre offset and are modelled accordingly. We estimate the flux ratio of $\mu$~Her Ab to $\mu$~Her Aa in the \textit{Gaia} band as $f_2/f_1 \approx 1/1000$, based on the measured magnitude differences from \cite{Roberts2016muHer}. Using this flux ratio, we compute $\Delta\eta$ for each epoch from the \textit{Gaia} GOST epochs and scan angles. Only four of the 64 modelled \textit{Gaia} observations are affected by light blending, and even in those cases the induced shift is only of order $0.03\,\mathrm{mas}$.

\subsection{Tangent-plane projection}  
When comparing absolute positions from different catalogues, we are really comparing two directions on the celestial sphere. These positions are reported in equatorial co-ordinates \((\alpha,\delta)\), but our model operates in a local Cartesian system---the tangent plane---centred on the \textsc{Hipparcos} reference position. Thus, taking a simple difference such as \(\alpha_{\rm G}-\alpha_{\rm H}\) implicitly assumes that curved-sky co-ordinates can be treated as flat. For small separations, this approximation is excellent. However, because $\mu$~Her has a large proper motion and the \textsc{Hipparcos}–\textit{Gaia} baseline spans $24.75$ years, the curvature of the sky becomes non-negligible. We therefore transform all non-\textsc{Hipparcos} catalogue positions into the tangent plane using the standard gnomonic projection \citep{Murray1983}. This effectively removes the curvature of the celestial sphere, allowing the corrected positions to be treated linearly in our models. For $\mu$~Her, this projection introduces a correction to positions of approximately $0.38\,\mathrm{mas}$ over the \textsc{Hipparcos}–\textit{Gaia} baseline.

\subsection{Perspective effects due to radial velocity}  
$\mu$~Her is approaching us with a CM radial velocity of around $-16.3\,\mathrm{km\,s}^{-1}$. As the system moves closer, its parallax increases. Additionally, its apparent proper motion grows too, since unchanged space motion projects differently as perspective changes. These effects are known as perspective acceleration \citep{Lindegren2021b}. Across the $24.75$-year \textsc{Hipparcos}–\textit{Gaia} baseline, the radial velocity changes the system parallax by ${\sim}6\,\mathrm{\mu as}$ and the total proper motion by ${\sim}0.082\,\mathrm{mas\,yr^{-1}}$. Accumulated over the baseline, the difference in proper motion displaces the system by ${\sim}1.5\,\mathrm{mas}$. In comparison with \textit{Gaia} DR$3$ uncertainties, the effect is negligible for parallax, but comparable to the errors in proper motion and clearly significant for positions (see Table~\ref{tab:reference_solutions}). Analogous to the deprojection procedure, we correct the catalogue positions and proper motions for perspective effects once, using a reference radial velocity of $-16.3\,\mathrm{km\,s^{-1}}$. This adjustment removes the geometric distortion caused by the system’s changing line-of-sight distance, without introducing additional parameters into the model itself.

\subsection{\textit{Gaia}–\textsc{Hipparcos} frame rotation}  
Although both \textsc{Hipparcos} and \textit{Gaia} are tied to the ICRS, their reference frames are not perfectly aligned. Small residual rotations and spins exist between them. Following the prescription of \citet{Fabricius2021}, we therefore rotate \textit{Gaia} astrometry into the \textsc{Hipparcos} frame before combining the datasets. In practice, this amounts to a global rigid transformation of the \textit{Gaia} 5p solution, ensuring that both catalogues describe the same inertial system. For DR$3$, the frame-rotation correction corresponds to about ${\sim}0.20\,\mathrm{mas\,yr^{-1}}$ in total proper motion and ${\sim}5.1\,\mathrm{mas}$ in position. The \citet{Munn2022} positions are anchored to the \textit{Gaia}~DR3 reference frame, and we therefore apply the same frame-rotation correction to these measurements before incorporating them into our model.

\section{Broadening function and corner plots}
\label{sec:appendixB}

Broadening function of the BC component is shown here along with corner plots from our MCMC sampler.

\begin{figure}
    \resizebox{\hsize}{!}{\includegraphics{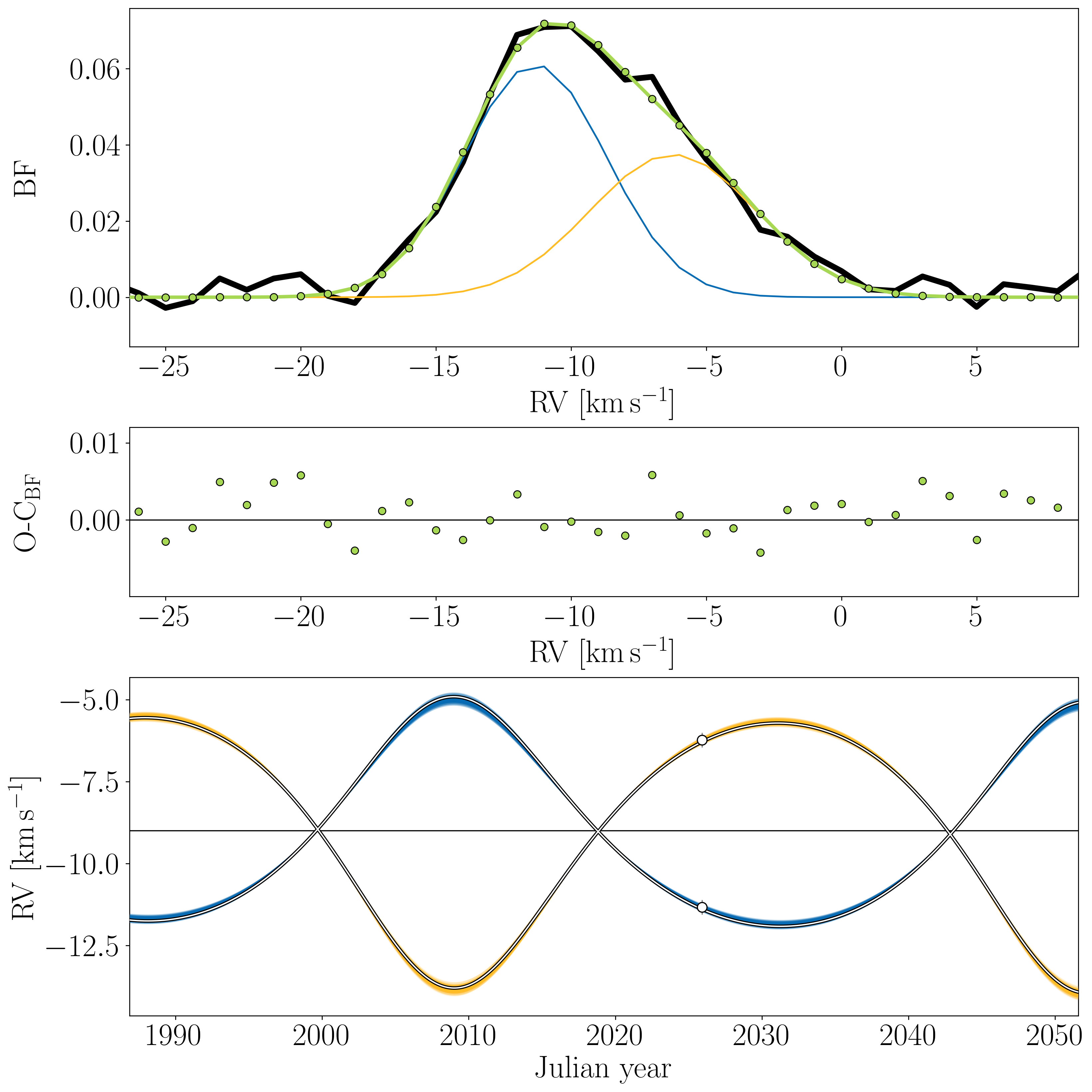}}
    \caption{Broadening function of $\mu$~Her~BC. \emph{Top:} Broadening function of $\mu$~Her~BC from our single December 2025 epoch. The best-fitting two-Gaussian model is overplotted (blue and yellow components; combined profile in green). \emph{Middle:} Residuals between the broadening function and the model. \emph{Bottom:} Radial-velocity curve of the BC subsystem. The two velocities extracted from the broadening function are indicated.}
    \label{fig:BF}
\end{figure}

\begin{figure*}
    \centering
    \includegraphics[width=\textwidth]{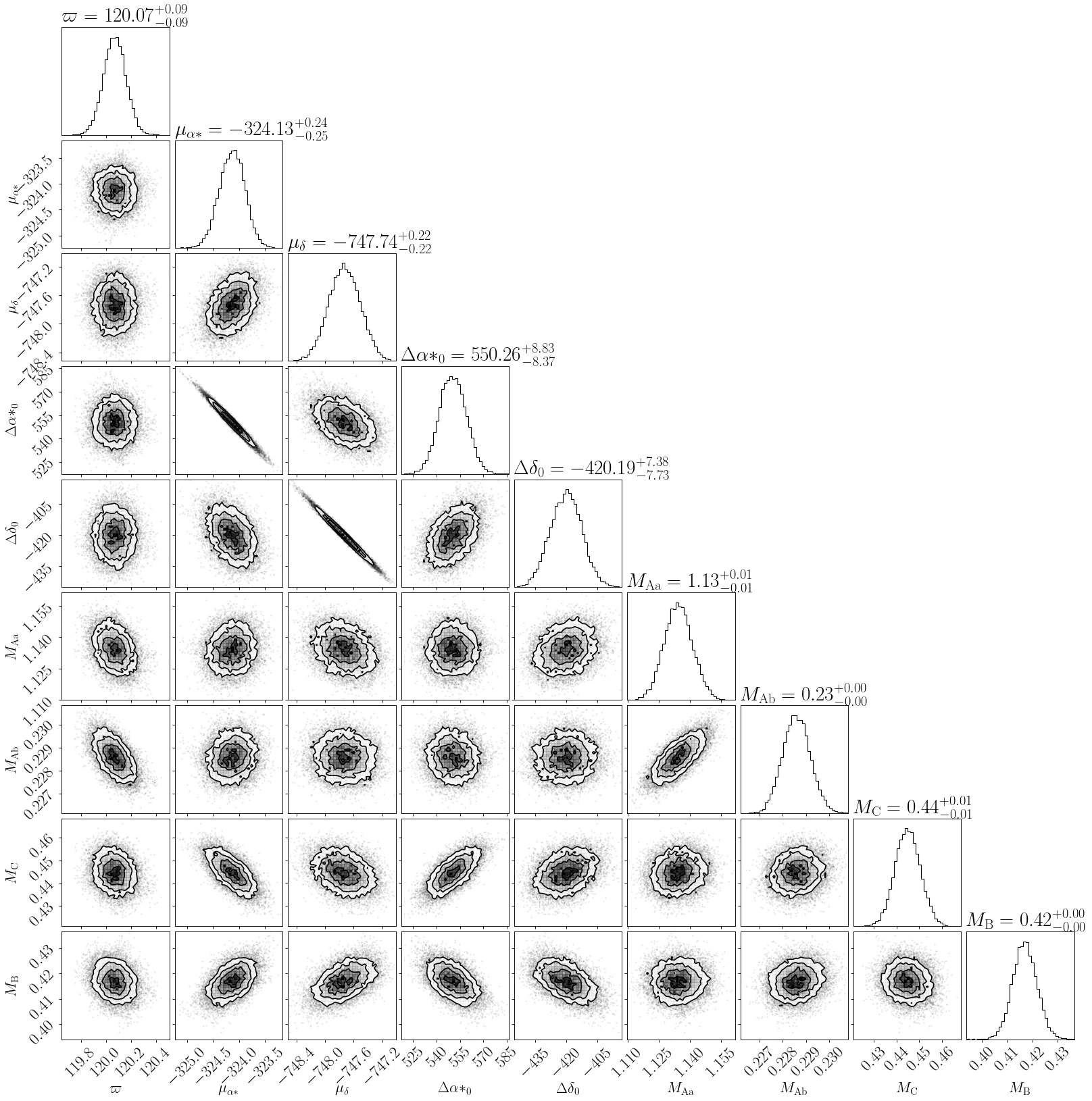}
    \caption{Corner plot showing the two-dimensional posterior distributions for a subset of parameters from our full joint fit. The panels display the astrometric parameters together with the derived masses of the four components.}
    \label{fig:corner_Astrometry}
\end{figure*}

\begin{figure*}
    \centering
    \includegraphics[width=\textwidth]{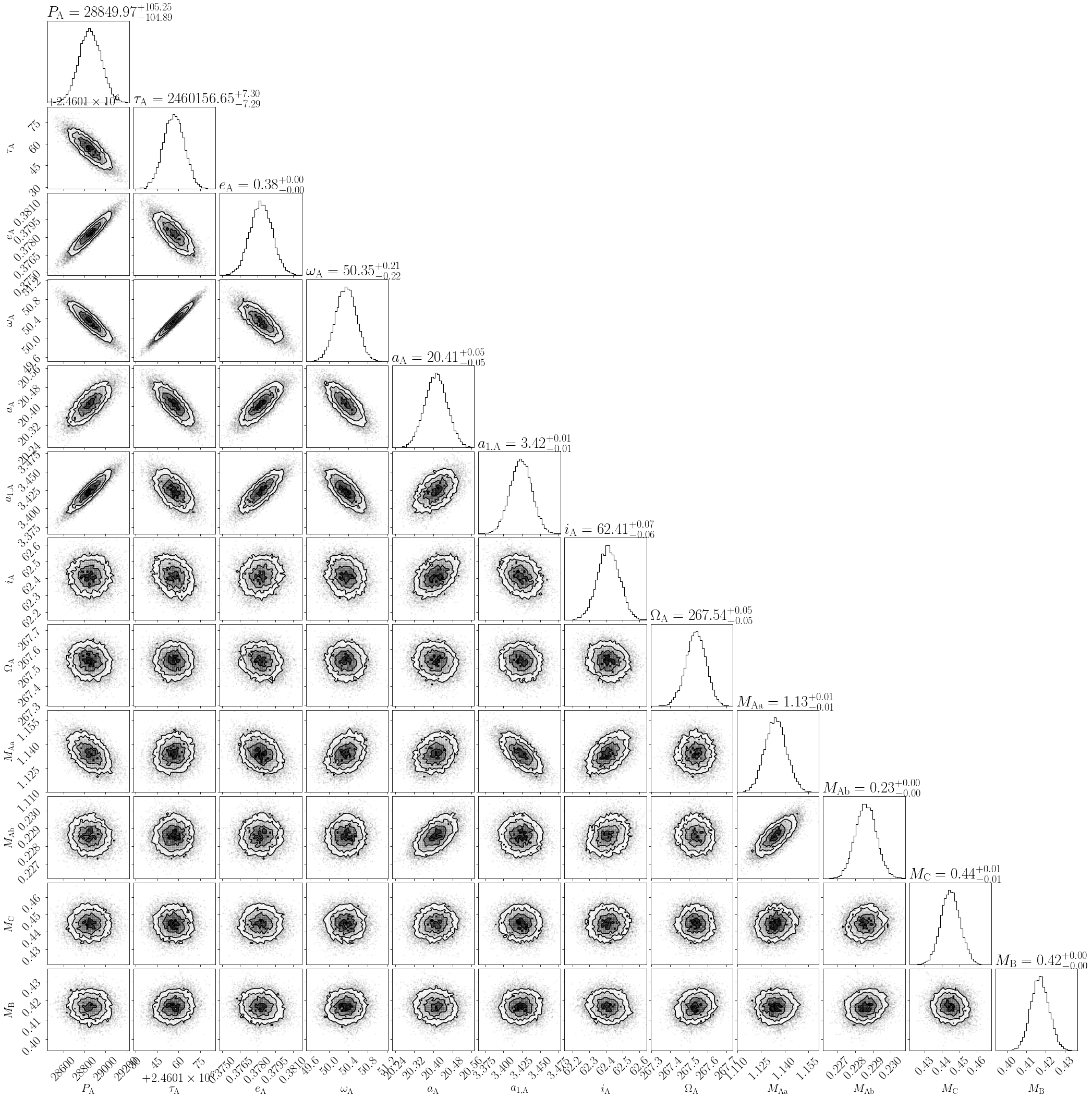}
    \caption{Corner plot of subsystem A's orbital parameters. The figure format is as the same as Fig.~\ref{fig:corner_Astrometry}.}
    \label{fig:corner_inner}
\end{figure*}

\begin{figure*}
    \centering
    \includegraphics[width=\textwidth]{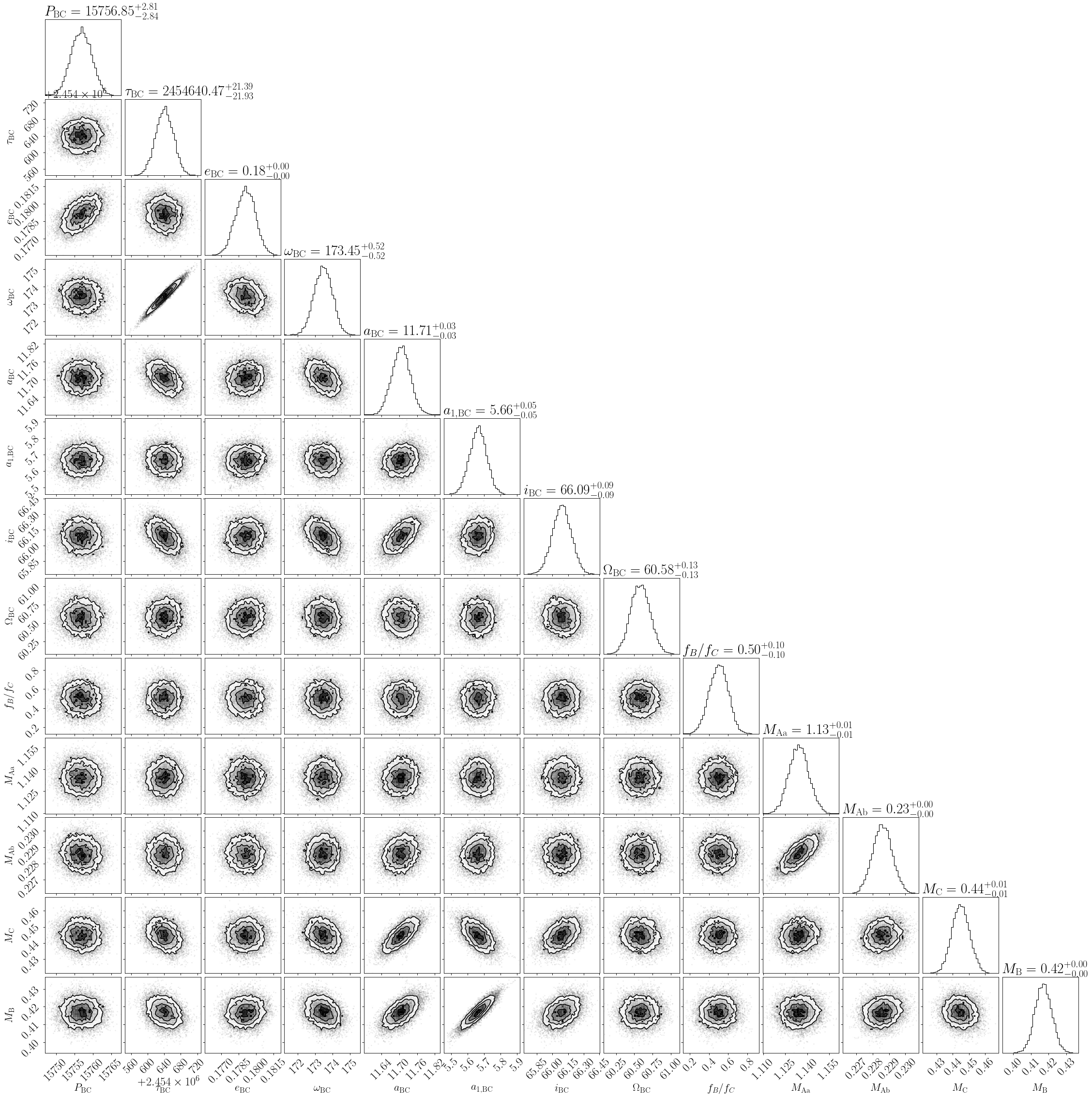}
    \caption{Corner plot of subsystem BC's orbital parameters. The figure format is as the same as Fig.~\ref{fig:corner_Astrometry}.}
    \label{fig:corner_BC}
\end{figure*}

\begin{figure*}
    \centering
    \includegraphics[width=\textwidth]{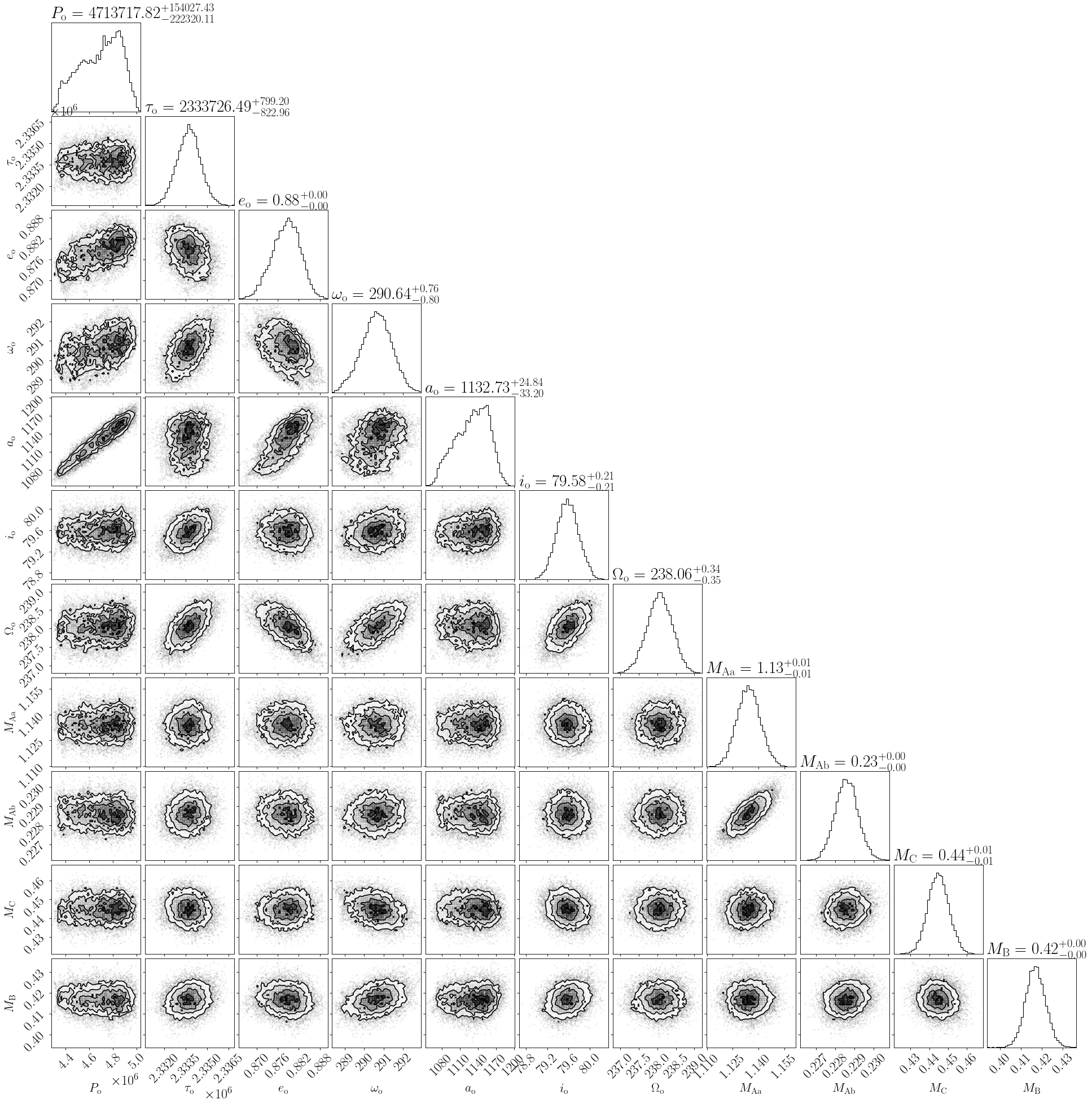}
    \caption{Corner plot of subsystem A--BC's orbital parameters. The figure format is as the same as Fig.~\ref{fig:corner_Astrometry}.}
    \label{fig:corner_outer}
\end{figure*}

\end{appendix}

\end{document}